\documentclass[article,nojss,shortnames]{jss}

\usepackage{thumbpdf,lmodern}
\usepackage{amsfonts,amstext,amsmath,amssymb,amsthm}
\usepackage{accents}
\usepackage{color}
\usepackage{rotating}
\usepackage{verbatim}
\usepackage{rotfloat}
\usepackage[nolists]{endfloat}
\DeclareDelayedFloatFlavor{sidewaystable}{table}
\DeclareDelayedFloatFlavor{sidewaysfigure}{figure}

\newcommand{\compact}{1}

\newcommand{\NEW}[1]{{#1}}

\newcommand{\rZ}{Z}
\newcommand{\rY}{Y}
\newcommand{\rX}{\mX}
\newcommand{\rU}{\mU}
\newcommand{\ru}{\uvec}
\newcommand{\rz}{z}
\newcommand{\ry}{y}
\newcommand{\rx}{\xvec}

\newcommand{\samZ}{\RR}
\newcommand{\samY}{\mathcal{Y}}
\newcommand{\samX}{\mathcal{X}}

\newcommand{\pZ}{F_\rZ}
\newcommand{\pY}{F_\rY}

\newcommand{\pSL}{F_{\SL}}
\newcommand{\pMEV}{F_{\MEV}}

\newcommand{\ProbYx}{\Prob_{\rY \mid \rX = \rx}}

\newcommand{\qZ}{F^{-1}_\rZ}

\newcommand{\dZ}{f_\rZ}
\newcommand{\dY}{f_\rY}

\newcommand{\h}{h}
\newcommand{\s}{\svec}

\newcommand{\hs}{\mathcal{H}}
\newcommand{\basisy}{\avec}
\newcommand{\bern}[1]{\avec_{\text{Bs},#1}}

\newcommand{\lik}{\mathcal{L}}
\newcommand{\parm}{\varthetavec}
\newcommand{\eparm}{\vartheta}
\newcommand{\dimparm}{P}

\newcommand{\shiftparm}{\betavec}

\newcommand{\reg}{r}

\newcommand{\ie}{i.e.,\ }
\newcommand{\eg}{e.g.,\ }

\renewcommand{\Prob}{\mathbb{P}}
\newcommand{\Ex}{\mathbb{E}}
\newcommand{\V}{\mathbb{V}}
\newcommand{\RR}{\mathbb{R}}

\usepackage{dsfont}

 \DeclareMathOperator*{\argmax}{{arg\,max}}

 \DeclareMathOperator{\ND}{N}

 \DeclareMathOperator{\SL}{SL}
 \DeclareMathOperator{\MEV}{MEV}

\def \avec {\text{\boldmath$a$}}

\def \svec {\text{\boldmath$s$}}    
    \def \mT {\text{\boldmath$T$}}
\def \uvec {\text{\boldmath$u$}}    \def \mU {\text{\boldmath$U$}}

\def \xvec {\text{\boldmath$x$}}    \def \mX {\text{\boldmath$X$}}

\def \betavec         {\text{\boldmath$\beta$}}

\def \varthetavec     {\text{\boldmath$\vartheta$}}

\newcommand{\ubar}[1]{\underaccent{\bar}{#1}}

\author{Torsten Hothorn \\ Universit\"at Z\"urich \\
   \And Achim Zeileis \\ Universit\"at Innsbruck}
\Plainauthor{Torsten Hothorn, Achim Zeileis}

\title{Transformation Forests} 

\Abstract{

Regression models for supervised learning problems with a continuous target
are commonly understood as models for the conditional mean of the target
given predictors.  This notion is simple and therefore appealing for
interpretation and visualisation.  Information about the whole underlying
conditional distribution is, however, not available from these models.  A
more general understanding of regression models as models for conditional
distributions allows much broader inference from such models, for example
the computation of prediction intervals.  Several random forest-type
algorithms aim at estimating conditional distributions, most prominently
quantile regression forests (Meins\-hausen, 2006, JMLR).  We propose a novel
approach based on a parametric family of distributions characterised by
their transformation function.  A dedicated novel ``transformation tree''
algorithm able to detect distributional changes is developed.  Based on
these transformation trees, we introduce ``transformation forests'' as an
adaptive local likelihood estimator of conditional distribution functions. 
The resulting models are fully parametric yet very general and allow broad
inference procedures, such as the model-based bootstrap, to be applied in a
straightforward way.
}


\Address{
  Torsten Hothorn\\
  Institut f\"ur Epidemiologie, Biostatistik und Pr\"avention \\
  Universit\"at Z\"urich \\
  Hirschengraben 84\\
  CH-8001 Z\"urich, Switzerland \\
  \texttt{Torsten.Hothorn@uzh.ch} \\

  Achim Zeileis \\
  Department of Statistics \\
  Faculty of Economics and Statistics \\
  Universit\"at Innsbruck \\
  Universit\"atsstra{\ss}e 15\\
  A-6020 Innsbruck, Austria
}

\begin{document}

\section{Introduction}

Supervised machine learning plays an important role in many prediction
problems.  Based on a learning sample consisting of $N$ pairs of target
value $\ry$ and predictors $\rx$, one learns a rule $\reg$ that predicts the
status of some unseen $\rY$ via $\reg(\rx)$ when only information about $\rx$
is available.  Both the machine learning and statistics communities
differentiate between ``classification problems'', where the target $\rY$ is
a class label, and ``regression problems'' with conceptually continuous
target observations $\ry$.  In binary classification problems with $\rY \in
\{0, 1\}$ the focus is on rules $\reg$ for the conditional probability of $\rY$ being $1$
given $\rx$, more formally $\Prob(\rY = 1 \mid \rX = \rx) = \reg(\rx)$.  Such a
classification rule $\reg$ is probabilistic in the sense that one cannot only
predict the most probable class label but also assess the corresponding
probability.  This additional information is extremely valuable because it
allows an assessment of the rules' $\reg$ uncertainty about its prediction.  It
is much harder to obtain such an assessment of uncertainty from most
contemporary regression models, because the rule (or ``regression
function'') $\reg$ typically describes the conditional expectation $\Ex(\rY
\mid \rX = \rx) = \reg(\rx)$ but not the full predictive distribution of $\rY$
given $\rx$.  Thus, the prediction $\reg(\rx)$ only contributes information
about the mean of some unseen target $\rY$ but tells us nothing about other
characteristics of its distribution.  Without making additional restrictive
assumptions, for example constant variances in normal distributions, the
derivation of probabilistic statements from the regression function $\reg$
alone is impossible.

\begin{figure}[t!]
\begin{center}
\includegraphics{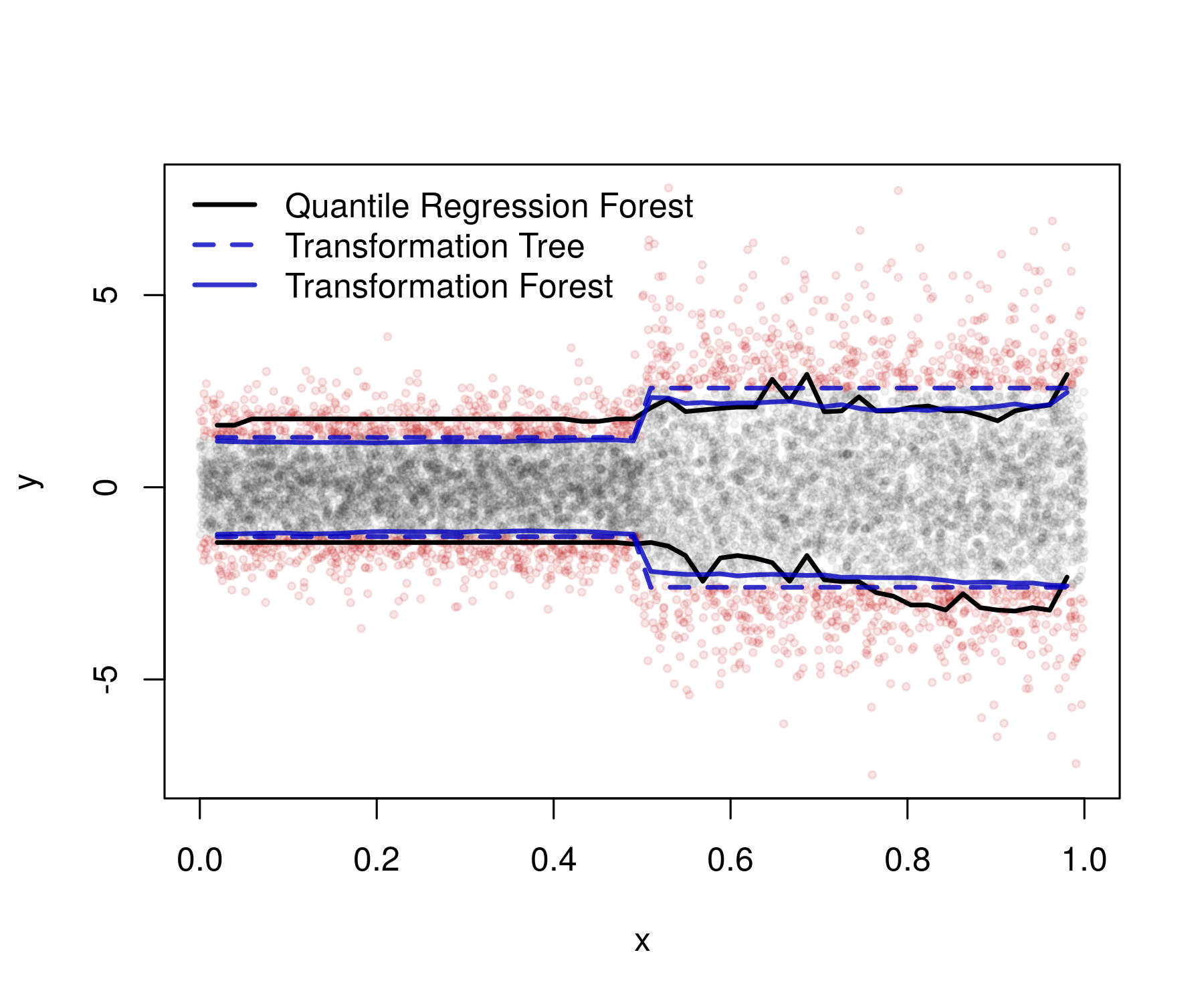}
\caption{Empirical Illustration.  Conditional on a uniform predictor $x$, the
distribution $\rY \sim \ND(0, (1 + I(x > .5))^2)$ features a
variance split at $.5$ for $10{,}000$ observations (points outside the
conditional $10\%$ and $90\%$ quantile are in red).  The black solid line
depicts estimated conditional $10\%$ and $90\%$ quantiles obtained from
quantile regression forests, the blue lines correspond to quantiles
estimated by transformation trees (dashed) and transformation forests (solid)
with non-linear transformation function parameterised via a Bernstein
polynomial of order five.
\label{fig:exIntro}}
\end{center}
\end{figure}

Contemporary random forest-type algorithms also strongly rely on the notion of
regression functions~$\reg$ describing the conditional mean $\Ex(\rY \mid \rX =
\rx)$ only \citep[for
example][]{Biau_Devroye_Lugosi_2008,Biau_2012,Scornet_Biau_Vert_2015},
although the first random forest-type algorithm for the estimation of
conditional distribution functions was published more than a decade ago
\citep[``bagging survival trees'',][]{Hothorn_Lausen_Benner_2004}.  A similar
approach was later developed independently by \cite{Meinshausen_2006} in his
``quantile regression forests''.  \NEW{In contrast to a mean aggregation
of cumulative hazard functions \citep{Ishwaran_Kogalur_Blackstone_2008} or densities
\citep{Criminisi_Shotton_Konukoglu_2011}, bagging survival trees and
quantile regression forests} are based on ``nearest neighbour
weights''. \NEW{We borrow this term from \cite{Lin_Jeon_2006}, where these weights 
were theoretically studied for the estimation of conditional means.}
The core idea is to obtain a ``distance'' measure based on the number of
times a pair of observations is assigned to the same terminal node in the
different trees of the forest.
Similar observations have a high probability of ending up in the same
terminal node whereas this probability is low for quite different
observations.  Then, the prediction for predictor values $\rx$ (either new or observed)
is simply obtained as a weighted empirical distribution function (or Kaplan-Meier estimator in the
context of right-censored target values) where those observations from the learning
sample similar (or dissimilar) to $\rx$ in the forest receive high (or low/zero)
weights, respectively.  Although this aggregation procedure in the aforementioned
algorithms is suitable for estimating predictive distributions, the
underlying trees are not.  The reason is that the ANOVA- or log-rank-type
split procedures commonly applied are not able to deal with distributions in
a general sense.  Consequently, the splits favour the detection of changes in the mean -- or
have power against proportional hazards alternatives in survival trees. However,
in general, they have very low power for detecting other patterns of heterogeneity
(e.g., changes in variance) even if these can be explained by the
predictor variables.  A simple toy example illustrating this problem is
given in Figure~\ref{fig:exIntro}.  Here, the target's conditional normal
distribution has a variance split at value $.5$ of a uniform $[0,1]$ predictor.  We
fitted a quantile regression forest \citep{Meinshausen_2006,
pkg:quantregForest} to the $10{,}000$ observations depicted in the figure along with
ten additional independent uniformly distributed non-informative predictors
(using $100$ trees without random variable selection; see Appendix~``Computational Details'').  
The true conditional $10\%$
and $90\%$ quantiles are not approximated very well by the quantile
regression forest.  In particular, the split at $.5$ does not play an
important role in this model.  Thus, although such an abrupt change in the
distribution can be represented by a binary tree, the traditional ANOVA
split criterion employed here was not able to detect this split.

To improve upon quantile regression forests and similar procedures in
situations where changes in moments beyond the mean are important, we
propose ``transformation forests'' for the
estimation and prediction of conditional distributions for $\rY$ given
predictor variables $\rx$ and proceed in three steps.  We first suggest to
understand forests as adaptive local likelihood estimators
\NEW{\citep[see][for a discussion of the special case of local linear regression]{Bloniarz_Wu_Yu_2016}.}  
Second, we recap the most important features of the flexible and
computationally attractive ``transformation family'' of distributions
\citep{Hothorn_Kneib_Buehlmann_2014,Hothorn_Moest_Buehlmann_2016} which
includes a variety of distribution families.
Finally, we adapt the core ideas of ``model-based
recursive partitioning'' \NEW{\citep[][who also provide a review of
earlier developments in this field]{Zeileis_Hothorn_Hornik_2008}} to this transformation
family and introduce
novel algorithms for ``transformation trees'' and ``transformation forests''
for the estimation of conditional distribution functions which potentially
vary in the mean and also in higher moments as a function of predictor
variables~$\rx$.  In our small example in Figure~\ref{fig:exIntro}, these novel
\NEW{transformation trees and forests} were able to recover the true conditional 
distributions much more precisely than quantile regression forests.

Owing to the fully parametric nature of the predictive distributions that
can be obtained from these novel methods, model inference procedures, such
as variable importances, independence tests or model-based resampling, can
be formulated in a very general and straightforward way (Section~\ref{sec:inference}).  
\NEW{Some remarks on asymptotic properties are given in Section~\ref{sec:theory}.
The performance of transformation trees and forests is evaluated empirically on
four artificial data generating processes and on survey data for body mass
indices from Switzerland in Section~\ref{sec:empeval}.  Details of
the variable and split selection procedure in transformation trees as well
as the corresponding theoretical complexity and empirical timings 
are discussed in Section~\ref{sec:complexity}.}

\section{Adaptive Local Likelihood Trees and Forests} \label{sec:lik}

We first deal with the unconditional distribution
$\Prob_\rY$ of a \NEW{target} random variable $\rY \in \samY$
and we restrict our attention to a specific
probability model defined by the parametric family of distributions
\begin{eqnarray*}
\Prob_{\rY, \Theta} = \{ \Prob_{\rY, \parm} \mid \parm \in \Theta \}
\end{eqnarray*}
with parameters $\parm$ and parameter space $\Theta \subseteq \RR^\dimparm$. 
With predictors $\rX = (X_1, \dots, X_J) \in \samX$ from some predictor sample space
$\samX$, our main interest is in the conditional distribution $\ProbYx$ and
we assume that this conditional distribution is a member of the
family of distributions introduced above, \ie we assume that a parameter
$\parm(\rx) \in \Theta$ exists such that $\ProbYx = \Prob_{\rY,
\parm(\rx)}$.  We call $\parm: \samX \rightarrow \Theta$ the ``conditional
parameter function'' and the task of estimating the conditional
distributions $\ProbYx$ for all $\rx$ reduces to the problem of estimating
this conditional parameter function.

From the probability model $\Prob_{\rY, \Theta}$ we can derive the
log-likelihood contribution $\ell_i: \Theta \rightarrow \RR$ 
for each of $N$ independent observations $(\ry_i, \rx_i)$ \NEW{from the learning
sample for} $i = 1, \dots, N$.
We propose and study a novel random forest-type estimator 
$\parm^N_\text{Forest}$ of the conditional parameter function $\parm$ in the 
class of adaptive local likelihood estimators of the form
\begin{eqnarray} \label{fm:locLik}
\hat{\parm}^N(\rx) := \argmax_{\parm \in \Theta} \sum_{i = 1}^N w_i^N(\rx)
\ell_i(\parm); \quad \rx \in \samX
\end{eqnarray}
where $w_i^N: \samX \rightarrow \RR^+$ is the ``conditional weight
function'' for observation $i$ given a specific configuration $\rx$
of the predictor variables (which may correspond to an observation from the
learning sample or to new data).  This weight measures the
similarity of the two distributions $\Prob_{\rY \mid \rX = \rx_i}$ and
$\Prob_{\rY \mid \rX = \rx}$ under the probability model $\Prob_{\rY,
\Theta}$.  The main idea is to obtain a large weight for observations $i$
which are ``close'' to $\rx$ in light of the model and essentially zero in
the opposite case.  The superscript $N$ indicates that the weight function
may depend on the learning sample, and in fact the choice of the weight
function $w_i^N$ is crucial in what follows.

Local likelihood estimation goes back to \cite{Brillinger_1977} in a comment
to \cite{Stone_1977} and was the topic of Robert Tibshirani's PhD thesis,
published in \cite{Tibshirani_Hastie_1987}.  Early regression models in this
class were based on the idea of fitting polynomial models locally within a
fixed smoothing window.  Adaptivity of the weights refers to an
$\rx$-dependent, non-constant smoothing window, \ie different weighing
schemes are applied in different parts of the predictor sample space
$\samX$.  An overview of local likelihood procedures was published by
\cite{Loader_1999}. Subsequently, we illustrate how classical
maximum likelihood estimators, model-based trees, and model-based forests
can be embedded in this general framework by choosing suitable conditional
weight functions and plugging these into~(\ref{fm:locLik}).

The unconditional maximum likelihood estimator is based on unit weights 
$w_{\text{ML},i}^N :\equiv 1$ not depending on $\rx$, \ie all observations
in the learning sample are considered to be equally ``close''; thus
\begin{eqnarray*}
\hat{\parm}_{\text{ML}}^N :=  \argmax_{\parm \in \Theta} \sum_{i = 1}^N \ell_i(\parm).
\end{eqnarray*}
In contrast, model-based trees can adapt to the learning sample by employing
rectangular splits to define a partition $\samX = \bigcup\limits_{b = 1, \dots,
B}^\bullet \mathcal{B}_b$ of the predictor sample space. Each of the $B$ cells
then contains a different local unconditional model. More precisely, the
conditional weight function $w_{\text{Tree},i}^N$ is simply an indicator for
$\rx_i$ and $\rx$ being elements of the same terminal node so that only
observations in the same terminal node are considered to be ``close''.
The weight and parameter functions are
\begin{align}
\begin{split} \label{fm:treeweights}
w_{\text{Tree},i}^N(\rx) := \sum_{b = 1}^B I(\rx \in \mathcal{B}_b \wedge \rx_i \in \mathcal{B}_b) \\
\hat{\parm}_{\text{Tree}}^N(\rx) :=  \argmax_{\parm \in \Theta} \sum_{i = 1}^N w_{\text{Tree},i}^N(\rx) \ell_i(\parm).
\end{split}
\end{align}
Thus, this essentially just picks the parameter estimate from the $b$-th
terminal node which is associated with cell $\mathcal{B}_b$
\begin{eqnarray*}
\hat{\parm}_b^N = \argmax_{\parm \in \Theta} \sum_{i = 1}^N I(\rx_i \in
\mathcal{B}_b) \ell_i(\parm),
\end{eqnarray*}
along with the corresponding conditional distribution $\Prob_{\rY, \hat{\parm}_b^N}$. Model-based
recursive partitioning \citep[MOB,][]{Zeileis_Hothorn_Hornik_2008} is one
representative of such a tree-structured approach.

A forest of $T$ trees is associated with partitions 
$\samX = \bigcup\limits_{b = 1, \dots, B_t}^\bullet \mathcal{B}_{tb}$ for
$t = 1, \dots, T$. The $b$-th terminal node of the $t$-th tree contains the 
parameter estimate $\hat{\parm}_{tb}^N$ and the $t$-th tree defines the  
conditional parameter function $\hat{\parm}_{\text{Tree}, t}^N(\rx)$.
We define the forest conditional parameter function via ``nearest neighbour'' 
forest weights
\begin{align} \label{fm:RFweights}
\begin{split}
w_{\text{Forest}, i}^N(\rx) := \sum_{t = 1}^T \sum_{b = 1}^{B_t} I(\rx \in \mathcal{B}_{tb}
\wedge \rx_i \in \mathcal{B}_{tb}) \\
\hat{\parm}_{\text{Forest}}^N(\rx) := \argmax_{\parm \in \Theta} \sum_{i = 1}^N w_{\text{Forest}, i}^N(\rx)
\ell_i(\parm).  
\end{split}
\end{align}
The conditional weight function $w_{\text{Forest}, i}^N$ counts how many times
$\rx_i$ and $\rx$ are
element of the same terminal node in each of the $T$ trees, \ie captures
how ``close'' the observations are on average across the trees in the forest.
\cite{Hothorn_Lausen_Benner_2004} first suggested these weights 
for the aggregation of $T$ survival trees.  The
same weights have later been used by \cite{Lin_Jeon_2006} \NEW{for estimating
conditional means, by \cite{Meinshausen_2006} for estimating conditional 
quantiles and by \cite{Bloniarz_Wu_Yu_2016} for estimating local linear
models.} An
``out-of-bag'' version only counts the contribution of the $t$-th tree for
observation $i$ when $i$ was not used for fitting the $t$-th tree.  

Forests relying on the aggregation scheme (\ref{fm:RFweights}) model the conditional
distribution $\ProbYx$ for some configuration $\rx$ of the
predictors as $\Prob_{\rY, \hat{\parm}_{\text{Forest}}^N(\rx)} \in
\Prob_{\rY, \Theta}$.  In this sense, such a forest 
is a fully specified parametric model with (in-bag or out-of-bag) log-likelihood
\begin{eqnarray*}
\sum_{i = 1}^N \ell_i\left(\hat{\parm}_{\text{Forest}}^N(\rx_i)\right)
\end{eqnarray*}
allowing a broad range of model inference procedures to be directly applied
as discussed in Section~\ref{sec:inference}.  Although this core idea seems
straightforward to implement, we unfortunately cannot pick our favourite
tree-growing algorithm and mix it with some parametric model as two critical
problems remain to be addressed in this paper.  First, most of the standard
tree-growing algorithms are not ready to be used for finding the underlying
partitions because their variable and split selection procedures have poor
power for detecting distributional changes which are not linked to changes
in the mean as was illustrated by the simple toy example presented in the
introduction.  Therefore, a tailored tree-growing algorithm inspired by model-based
recursive partitioning also able to detect changes in higher moments is
introduced in Section~\ref{sec:trees}.  The second problem is associated
with the parametric families $\Prob_{\rY, \Theta}$.  Although, in principle,
all classical probability models are suitable in this general framework,
different parameterizations render unified presentation and especially
implementation burdensome.  We address this second problem by restricting
our implementation to a novel transformation family of distributions. 
Theoretically, this family contains all univariate probability distributions
$\Prob_\rY$ and practically close approximations thereof.  We highlight
important aspects of this family and the corresponding likelihood function 
in the next section.

\section{Transformation Models} \label{sec:trafo}

A transformation model $\Prob(\rY \le \ry) = \pY(\ry) = \pZ(\h(\ry))$ describes the
distribution function of $\rY$ by an unknown monotone increasing
transformation function $\h$ and some a priori chosen continuous distribution function
$\pZ$.  We use this framework because simple, \eg linear,
transformation
functions implement many of the classical parametric models whereas more
complex transformation functions provide similar flexibility as models
from the non-parametric world.  In addition, discrete and continuous targets, also under all
forms of random censoring and truncation, are handled in a unified way. As a
consequence, our corresponding ``transformation forests'' will be applicable
to a wide range of targets (discrete, continuous with or without censoring
and truncation, counts, survival times) with the option to gradually move
from simple to very flexible models for the conditional distribution
functions $\Prob_{\rY, \hat{\parm}_\text{Forest}^N(\rx)}$.

In more detail, let $\rZ \sim \Prob_\rZ$ denote an
absolutely continuous random variable with 
density, distribution, and quantile functions $\dZ$, $\pZ$ and $\qZ$,
respectively.  We furthermore assume $0 < \dZ(\rz) < \infty \, \forall \rz
\in \samZ$ for a log-concave
density $\dZ$ as well as the existence of the first two derivatives of the
density $\dZ(\rz)$ with respect to $\rz$, both derivatives shall be bounded. 
We do not allow any unknown parameters for this distribution.  Possible
choices include the standard normal, the standard logistic and the
\NEW{standard} minimum
extreme value distribution with distribution functions $\pZ(\rz) =
\Phi(\rz)$, $\pZ(\rz) = \pSL(\rz) = (1 + \exp(-\rz))^{-1}$ and $\pZ(\rz) =
\pMEV(\rz) = 1 -\exp(-\exp(\rz))$, respectively.

Let $\hs = \{\h: \samY \rightarrow \samZ \mid 
\h(\ry_1) < \h(\ry_2) \, \forall \ry_1 < \ry_2 \in \samY\}$ denote the
space of all strictly monotone transformation functions. 
With the transformation function $\h$ we can write $\pY$ as
$\pY(\ry \mid \h) = \pZ(\h(\ry)) \, \forall \ry \in \samY$ with
density $\dY(\ry \mid \h)$ and there exists a
unique transformation function $\h = \pZ^{-1} \circ \pY$ for all distribution functions 
$\pY$ \citep{Hothorn_Moest_Buehlmann_2016}. A convenient
feature of characterising the distribution of $\rY$ by means of the
transformation function $\h$ is that the likelihood for arbitrary 
measurements can be written and implemented in an extremely compact form.

\NEW{For a given transformation function $h$, the likelihood contribution of an
observation $\ry \in \RR$ is given by the corresponding density
\begin{eqnarray*}
\lik(\h \mid \rY = \ry) = \dZ(\h(\ry)) \h^\prime(\ry).
\end{eqnarray*}
The likelihood for intervals $(\ubar{\ry},\bar{\ry}] \subset \samY$ 
is, unlike in the above  ``exact continuous'' case, defined in terms of the
distribution function \citep{Lindsey_1996}, where one can differentiate
between three special cases:
\begin{eqnarray*}
\lik(\h \mid \rY \in (\ubar{\ry},\bar{\ry}]) = \left\{ \begin{array}{ll}
     \pZ(\h(\bar{\ry})) - \pZ(\h(\ubar{\ry})) & \ry \in
(\ubar{\ry},\bar{\ry}] \quad    \text{``interval-censored''} \\
     1 - \pZ(\h(\ubar{\ry})) & \ry \in (\ubar{\ry}, \infty) 
\quad \text{``right-censored''} \\
     \pZ(\h(\bar{\ry})) & \ry \in (-\infty, \bar{\ry}]\quad \text{``left-censored''}
\end{array} \right.
\end{eqnarray*}
For truncated observations in the interval $(\ry_l, \ry_r] \subset \samY$,
the above likelihood contribution has to be multiplied by the factor
$(\pZ\left(\h(\ry_r)\right) - \pZ\left(\h(\ry_l)\right))^{-1}$
when $\ry_l < \ubar{\ry} < \bar{\ry} \le \ry_r$.
A more detailed discussion of likelihood contributions to transformation models can
be found in \cite{Hothorn_Moest_Buehlmann_2016}.}

We parameterise the transformation function $\h(\ry)$ as a linear function
of its basis-transformed argument $\ry$ using a basis function $\basisy: \samY   
\rightarrow \RR^\dimparm$ such that $\h(\ry) = \basisy(\ry)^\top \parm,
\parm \in \RR^\dimparm$.  In the following, we will write $\h =
\basisy^\top \parm$ \NEW{and assume that the true unknown transformation function
is of this form.}
For continuous targets $\rY$ the parameterisation $\h(\ry) =
\basisy(\ry)^\top \parm$ needs to be smooth in $\ry$, so any polynomial or
spline basis is a suitable choice for $\basisy$.  For the empirical
experiments in Section~\ref{sec:empeval} we employed Bernstein polynomials
\citep[for an overview see][]{Farouki_2012} of order $M$ ($\dimparm = M +
1$) defined on the interval $[\ubar{\imath}, \bar{\imath}]$ with
\begin{eqnarray*}
\bern{M}(\ry) & = & (M + 1)^{-1}(f_{\text{Be}(1, M + 1)}(\tilde{\ry}),
\dots,
                            f_{\text{Be}(m, M - m + 1)}(\tilde{\ry}), \dots,
                            f_{\text{Be}(M + 1, 1)}(\tilde{\ry}))^\top \in
\RR^{M + 1} \\
\h(\ry) & = & \bern{M}(\ry)^\top \parm =
              \sum_{m = 0}^{M} \eparm_m f_{\text{Be}(m + 1, M - m +
1)}(\tilde{\ry}) / (M + 1) \\
\h^\prime(\ry) & = & \bern{M}^\prime(\ry)^\top \parm =
              \sum_{m = 0}^{M - 1} (\eparm_{m + 1} - \eparm_m)
f_{\text{Be}(m + 1, M - m)}(\tilde{\ry}) M /
((M + 1) (\bar{\imath} - \ubar{\imath}))
\end{eqnarray*}
where $\tilde{\ry} = (\ry -\ubar{\imath}) / (\bar{\imath} - \ubar{\imath})
\in [0, 1]$ and $f_{\text{Be}(m, M)}$ is the density of the Beta
distribution with parameters $m$ and $M$.  This choice is computationally
attractive because strict monotonicity can be formulated as a set of $M$
linear constraints on the parameters $\eparm_m < \eparm_{m + 1}$ for all $m
= 0, \dots, M$ \citep{Curtis_Ghosh_2011}.  

The distribution family 
\if0\compact
\begin{eqnarray*}
\Prob_{\rY, \Theta} = \{ \pZ \circ \basisy^\top\parm \mid \parm \in \Theta \}
\end{eqnarray*}
\else
$\Prob_{\rY, \Theta} = \{ \pZ \circ \basisy^\top\parm \mid \parm \in \Theta \}$
\fi
that transformation forests are based upon is called transformation family of distributions 
with parameter space $\Theta = \{\parm \in \RR^\dimparm \mid \basisy^\top\parm
\in \hs\}$ and transformation functions
$\basisy^\top\parm \in \hs$. 
\NEW{This family encompasses a wide variety of densities capturing different
locations and shapes (including scale and skewness), see
Figure~\ref{BMI-tree-plot} for an illustration of different body mass index
distributions.}
The log-likelihood contribution for an observation $\ry_i \in \RR$ is now the log-density
of the transformation model $\ell_i(\parm) = \log(\dZ(\basisy(\ry_i)^\top \parm)) + \log(\basisy^\prime(\ry_i)^\top \parm)$.

\section{Transformation Trees and Forests} \label{sec:trees}

Conceptually, the model-based recursive partitioning algorithm \citep{Zeileis_Hothorn_Hornik_2008}
for tree induction starts with the maximum likelihood estimator
$\hat{\parm}_\text{ML}^N$. Deviations from such a given model that can be
explained by parameter instabilities due to one or more of the predictors
are investigated based on the score contributions. The novel 
``transformation trees'' suggested here rely on the transformation family
$\Prob_{\rY, \Theta} = \{ \pZ \circ \basisy^\top\parm \mid \parm \in \Theta
\}$ whose score contributions $\s$
have relatively simple and generic forms. The score contribution of an ``exact continuous''
observation $\ry \in \RR$ from an absolutely continuous distribution is
given by
the
gradient of the log-density with respect to $\parm$
\if0\compact
\begin{eqnarray*}
\s(\parm \mid \rY = \ry) =
\frac{\partial \log(\dY(\ry \mid \parm))}{\partial \parm} & = &
\frac{\partial \log(\dZ(\basisy(\ry)^\top \parm))) +
\log({\basisy^\prime(\ry)}^\top \parm)}{\partial \parm} \nonumber \\
& = & \basisy(\ry) \frac{\dZ^\prime(\basisy(\ry)^\top
\parm)}{\dZ(\basisy(\ry)^\top \parm)}
    + \frac{\basisy^\prime(\ry)}{{\basisy^\prime(\ry)}^\top \parm}.
\label{f:s_exact}
\end{eqnarray*}
\else
\begin{eqnarray*}
\s(\parm \mid \rY = \ry) =
\basisy(\ry) \frac{\dZ^\prime(\basisy(\ry)^\top
\parm)}{\dZ(\basisy(\ry)^\top \parm)}
    + \frac{\basisy^\prime(\ry)}{{\basisy^\prime(\ry)}^\top \parm}.
\label{f:s_exact}
\end{eqnarray*}
\fi
For an interval-censored observation $(\ubar{\ry}, \bar{\ry}]$ the score contribution is
\if0\compact
\begin{eqnarray*}
\s(\parm \mid \rY \in (\ubar{\ry}, \bar{\ry}]) & = & \frac{\partial
\log(\lik(\basisy^\top \parm \mid \rY \in (\ubar{\ry},
\bar{\ry}]))}{\partial \parm} \nonumber \\
& = & \frac{\partial \log(\pZ(\basisy(\bar{\ry})^\top \parm) -
\pZ(\basisy(\ubar{\ry})^\top \parm))}{\partial \parm}  \nonumber \\
& = & \frac{\dZ(\basisy(\bar{\ry})^\top \parm)\basisy(\bar{\ry}) -
\dZ(\basisy(\ubar{\ry})^\top
\parm) \basisy(\ubar{\ry})}{\pZ(\basisy(\bar{\ry})^\top \parm) -
\pZ(\basisy(\ubar{\ry})^\top
\parm)}. \label{f:s_interval}
\end{eqnarray*}
\else
\begin{eqnarray*}
\s(\parm \mid \rY \in (\ubar{\ry}, \bar{\ry}]) =
\frac{\dZ(\basisy(\bar{\ry})^\top \parm)\basisy(\bar{\ry}) -
\dZ(\basisy(\ubar{\ry})^\top
\parm) \basisy(\ubar{\ry})}{\pZ(\basisy(\bar{\ry})^\top \parm) -
\pZ(\basisy(\ubar{\ry})^\top
\parm)}. \label{f:s_interval}
\end{eqnarray*}
\fi
Under truncation to the interval $(\ry_l, \ry_r] \subset \samY$, 
one needs to substract the term $\s(\parm \mid \rY \in (\ry_l, \ry_r])$ from the score function.

\NEW{With the transformation model and thus the likelihood and score
function being available, we start our tree induction with the global} model $\Prob_{\rY, \hat{\parm}_\text{ML}^N}$. The
hypothesis of all observations $i = 1, \dots, N$ coming from this model
can be written as the independence of the $\dimparm$-dimensional score contributions and all
predictors, \ie
\begin{eqnarray*}
H_0: \s(\hat{\parm}_\text{ML}^N \mid \rY ) \perp \rX.
\end{eqnarray*}
This hypothesis can be tested either using asymptotic M-fluctuation tests
\citep{Zeileis_Hothorn_Hornik_2008} or permutation tests 
\citep{Hothorn_Hornik_Zeileis_2006,Zeileis_Hothorn_2013} with appropriate
multiplicity adjustment depending on the number of predictors.
\NEW{Rejection of $H_0$ leads to the implementation of a binary split
in the predictor variable with \NEW{most significant} association to the score matrix;
algorithmic details are discussed in Section~\ref{sec:complexity}.
Unbiasedness of a model-based tree with respect to variable
selection is a consequence of splitting in the variable of highest
association to the scores where association is measured by the marginal
multiplicity-adjusted $p$-value \citep[for details
see][and Section~\ref{sec:complexity}]{Hothorn_Hornik_Zeileis_2006,Zeileis_Hothorn_Hornik_2008}.}
The procedure is recursively iterated until $H_0$
cannot be rejected. The result is a partition of the sample space
$\samX = \bigcup\limits_{b = 1, \dots, B}^\bullet \mathcal{B}_b$.

Based on the ``transformation trees'' introduced here, we construct a corresponding
random forest-type algorithm as follows.
A ``transformation forest'' is an ensemble of $T$ transformation trees fitted to 
subsamples of the learning sample and, optionally, a random selection of
candidate predictors available for splitting in each node of the
tree. The result is a set of $T$ partitions of the predictor sample space.
The transformation forest conditional parameter function 
is defined by its nearest neighbour forest weights~(\ref{fm:RFweights}).


The question arises how the order $M$ of the parameterisation of the
transformation function $\h$ via Bernstein polynomials 
affects the conditional distribution functions
$\Prob_{\rY, \hat{\parm}_{\text{Tree}}^N(\rx)}$ and $\Prob_{\rY,
\hat{\parm}_{\text{Forest}}^N(\rx)}$.  On the one hand, the basis $\bern{1}$
with $\pZ = \Phi$ only allows linear transformation functions of a standard
normal and thus our models for $\Prob_{\rY \mid \rX = \rx}$ are restricted
to the normal family, however, with potentially both mean and variance
depending on $\rx$ as the split criterion in transformation trees is
sensitive to changes in both mean and variance.  This most simple
parameterisation leads to transformation trees and forests from which both the conditional mean and the
conditional variance can be inferred. Using a higher order $M$
also allows modelling non-normal distributions. In the extreme case with $M
= N - 1$ the unconditional distribution function $\pZ(\bern{M}(\ry)^\top
\parm)$ interpolates the unconditional empirical cumulative distribution
function of the target.  With $M > 1$, the split criterion introduced in
this section is able to detect changes beyond the second moment and,
consequently, also higher moments of the conditional distributions $\Prob_{\rY \mid \rX =
\rx}$ may vary with $\rx$. An empirical comparison of transformation trees and forests
with linear ($M = 1$) and nonlinear ($M > 1$) transformation function
can be found in Section~\ref{sec:empeval}. \NEW{Additional empirical properties
of transformation models with larger values of $M$ are discussed
in \cite{Hothorn_2018}.}

\section{Transformation Forest Inference} \label{sec:inference}

In contrast to other random forest regression models,
a transformation forest is a fully-specified parametric model.  Thus, we can
derive all interesting model inference procedures from well-defined
probability models and do not need to fall back to heuristics.  Predictions
from transformation models are distributions $\Prob_{\rY,
\hat{\parm}_\text{Forest}^N(\rx)}$ and we can describe these on the scale of
the distribution, quantile, density, hazard, cumulative hazard, expectile,
and any other characterising functions.  By far not being comprehensive, we
introduce prediction intervals, a unified definition of permutation variable
importance, the model-based bootstrap and a test for global independence
in this section.

\subsection{Prediction Intervals and Outlier Detection}

For some yet unobserved target $\rY$ under predictors $\rx$,  
a two-sided $(1 - \alpha)$ prediction interval for $\rY \mid \rX = \rx$ and
some $\alpha \in (0, .5)$ can be obtained by 
numerical inversion of the conditional distribution $\Prob_{\rY,
\hat{\parm}_\text{Forest}^N(\rx)}$, for example via
\begin{eqnarray*}
\text{PI}_\alpha(\rx \mid \hat{\parm}_\text{Forest}^N) = 
  \left\{\ry \in \samY \mid \alpha / 2 < \Prob_{\rY, \hat{\parm}_\text{Forest}^N(\rx)}(\ry) \le 1 -
\alpha/2 \right\}
\end{eqnarray*}
with the property
\begin{eqnarray*}
\Prob_{\rY \mid \rX = \rx}\left(\text{PI}_\alpha(\rx \mid \parm)\right) = 1 - \alpha.
\end{eqnarray*}
The empirical level $\Prob_{\rY \mid \rX = \rx}(\text{PI}_\alpha(\rx \mid
\hat{\parm}_\text{Forest}^N))$ depends on how well the parameters
$\parm(\rx)$ are approximated by the forest estimate
$\hat{\parm}_\text{Forest}^N(\rx)$. If for some observation $(\ry_i,
\rx_i)$ the corresponding prediction interval $\text{PI}_\alpha(\rx_i \mid
\hat{\parm}_\text{Forest}^N)$ excludes $\ry_i$, one can (at level $\alpha$) 
suspect this observed target of being an outlier.

\subsection{Permutation Variable Importance} \label{sec:varimp}

The importance of a variable is defined as the amount of change in the risk
function when the association between one predictor variable and the
target is artificially broken. Permutation variable importances permute
one of the predictors at a time \NEW{\citep[and thus also break the association to the
remaining predictors, see][]{Strobl_Boulesteix_Kneib_2008}}. 
The risk function for transformation forests 
is the negative log-likelihood, thus a universally applicable formulation
of variable importance for all types of target distributions in
transformation forests is
\begin{eqnarray*}
\text{VI}(j) = T^{-1} \sum_{t = 1}^T \left(\sum_{i = 1}^N
-\ell_i\left(\hat{\parm}_{\text{Tree}, t}^N(\rx_i)\right) - \sum_{i = 1}^N
-\ell_i\left(\hat{\parm}_{\text{Tree}, t}^N(\rx_i^{(j)})\right)\right)
\end{eqnarray*}
where the $j$-th variable was permuted in $\rx_i^{(j)}$ for $i = 1, \dots,
N$.

\subsection{Model-Based Bootstrap} \label{sec:modelboot}

We \NEW{suggest} the model-based, or ``parametric'', bootstrap to assess the variability of the estimated
forest conditional parameter function $\hat{\parm}_{\text{Forest}}^N$ as
follows. First, we fit a transformation forest and sample
new target values $\tilde{\ry}_i \sim 
\Prob_{\rY, \hat{\parm}_\text{Forest}^N(\rx_i)}$
for each observation $i = 1, \dots, N$ 
from this transformation forest. For these $i = 1, \dots, N$
pairs of artificial targets and original predictors 
$(\tilde{\ry}_i, \rx_i)$, we refit the transformation forest. This procedure
of sampling and refitting is repeated $k = 1, \dots, K$ times. The resulting
$K$ conditional parameter functions $\hat{\parm}_{\text{Forest}, k}^N, k =
1, \dots, K$ are
a bootstrap sample from the distribution of conditional parameter functions assuming
the initial $\hat{\parm}_\text{Forest}^N$ was the true conditional parameter
function. \NEW{The bootstrap distribution of $\hat{\parm}_{\text{Forest},
k}^N(\rx)$ or functionals thereof can be used to study their variability or
to derive bootstrap confidence intervals \citep{Efron_Tibshirani_1993} for
parameters $\parm(\rx)
$ or other quantities, such as conditional quantiles.}

\subsection{Independence Likelihood-Ratio Test} \label{sec:LR}

The first question many researchers have is ``Is there any signal in my
data?'', or, in other words, is the target $\rY$ independent of all
predictors $\rX$? Classical tests, such as the $F$-test in a linear model
or multiplicity-adjusted univariate tests, have very low power against 
complex alternatives, \ie in situations where the impact of the predictors
is neither linear nor marginally visible. Because transformation forests
can potentially detect such structures, we propose a likelihood-ratio test
for the null $H_0: \rY \perp \rX$. This null hypothesis is identical to 
$H_0: \Prob_{\rY} = \Prob_{\rY \mid \rX = \rx} \forall \rx \in \samX$ and
reads
$H_0: \Prob_{\rY, \parm} = \Prob_{\rY, \parm(\rx)} \forall \rx \in \samX$,
or even simpler, $H_0: \parm(\rx) \equiv \parm$ for the class of models we
are studying. Under the null hypothesis, the 
unconditional maximum likelihood estimator $\hat{\parm}_\text{ML}^N$ would be optimal. It
therefore makes sense to compare the log-likelihoods of the unconditional
model with the log-likelihood of the transformation forest using the
log-likelihood ratio statistic
\begin{eqnarray*}
\text{logLR} = \sum_{i = 1}^N
\ell_i\left(\hat{\parm}_\text{Forest}^N(\rx_i)\right) -
\sum_{i = 1}^N \ell_i\left(\hat{\parm}_\text{ML}^N\right)
\end{eqnarray*}
Under $H_0$ we expect small differences and under the alternative we expect
to see larger log-likelihoods of the transformation forest. 
\NEW{The null distribution of such likelihood-ratio statistics 
is hard to assess analytically but can be easily approximated by the model-based 
bootstrap \citep[early references include][]{McLachlan_1987,Beran_1988}}.
We first estimate the unconditional model $\Prob_{\rY, \hat{\parm}_\text{ML}^N}$ and,
in a second step, draw $k = 1, \dots, K$ samples from this model $\Prob_{\rY,
\hat{\parm}_\text{ML}^N}$ of size $N$, \ie we sample from the unconditional
model, in this sense treating
$\hat{\parm}_\text{ML}^N$ as the ``true'' parameter. In the $k$-th sample the predictors 
are identical to the those in the learning sample and only the target values
are replaced. For each of these $k$ samples we refit the transformation
forest and obtain $\hat{\parm}_{\text{Forest}, k}^N(\rx_i)$. Based on this
model we compute the log-likelihood
ratio statistic
\begin{eqnarray*}
\text{logLR}_k = \sum_{i = 1}^N \ell_{i,k}\left(\hat{\parm}_{\text{Forest},
k}^N(\rx_i)\right) - 
\sum_{i = 1}^N \ell_{i,k}\left(\hat{\parm}_\text{ML}^N\right)
\end{eqnarray*}
where $\ell_{i,k}$ is the log-likelihood contribution by the $i$-th
observation from the $k$-th bootstrap sample. The $p$-value for $H_0$ is now
$K^{-1} \sum_k I(\text{logLR}_k > \text{logLR})$. \NEW{The size of this test in
finite samples depends on the performance of transformation forests under
$H_0$ and its power on the ability of transformation forests to detect
non-constant conditional parameter functions $\parm(\rx)$. Empirical
evidence for a moderate overfitting behaviour and a high power for detecting
distributional changes are reported in Section~\ref{sec:empeval}.}




\section{Theoretical Evaluation} \label{sec:theory}

The theoretical properties of random forest-type algorithms are a
contemporary research problem  and we refer to \cite{Biau_Scornet_2016}
for an overview. In this section we discuss how these developments relate to
the asymptotic behaviour of transformation trees and transformation forests.

For $\parm(\rx) \equiv \parm$ the maximum likelihood estimator ($w_i \equiv
1$) is consistent and asymptotically normal
\citep{Hothorn_Moest_Buehlmann_2016}.  In the non-parametric setup, \ie for
arbitrary distributions $\Prob_\rY$, \cite{Hothorn_Kneib_Buehlmann_2014}
provide consistency results in the class of conditional transformation
models.  Based on these results, consistency and normality of the local
likelihood estimator for an a priori known partition $\samX =
\bigcup\limits_{b = 1, \dots, B}^\bullet \mathcal{B}_b$
is guaranteed as long as the sample size tends to infinity
in all cells $b$.

If the partition (transformation trees) or the nearest neighbour weights
(transformation forests) are estimated from the data, established theoretical results on
random forests \NEW{\citep{Breiman_2004, Lin_Jeon_2006, Meinshausen_2006, Biau_Devroye_Lugosi_2008, 
Biau_2012, Scornet_Biau_Vert_2015}} provide a basis for the analysis of
transformation forests. \cite{Lin_Jeon_2006} first analysed random forests
for estimating conditional means with adaptive nearest
neighbours weights, where estimators for the conditional mean of the form
\begin{eqnarray*}
\hat{\Ex}_N(\rY \mid \rX = \rx) = \frac{\sum_{i = 1}^N w_{\text{Forest}, i}^N(\rx)
\rY_i}{\sum_{i = 1}^N w_{\text{Forest}, i}^N(\rx)}
\end{eqnarray*}
were shown to be consistent in non-adaptive random forests
\begin{eqnarray*}
\Ex_{\rY \mid \rX = \rx}\left(\Ex(\rY \mid \rX = \rx) - \hat{\Ex}_N(\rY
\mid \rX = \rx)\right)^2 \rightarrow 0
\end{eqnarray*}
as $N \rightarrow \infty$. \cite{Meinshausen_2006}
showed a Glivenko-Cantelli-type result for conditional distribution
functions
\begin{eqnarray} \label{fm:QRF}
\hat{\Prob}_N (\rY \le \ry \mid \rX = \rx) = \hat{\Ex}_N(I(\rY \le \ry) \mid \rX = \rx) =
\frac{\sum_{i = 1}^N w_{\text{randomForest}, i}^N(\rx) I(\rY_i \le \ry)}{\sum_{i =
1}^N w_{\text{randomForest}, i}^N(\rx)}
\end{eqnarray}
where the weights are obtained from Breiman and Cutler's original random forest
implementation \citep{Breiman_2001}.

In order to understand the applicability of these results to transformation
forests, we define the expected conditional log-likelihood given $\rx$ for a fixed set of parameters $\parm$ as
\begin{eqnarray*}
\ell(\parm \mid \rX = \rx) := \Ex_{\rY \mid \rX = \rx} \ell(\parm, \rY),
\end{eqnarray*}
where $\ell(\parm, \rY_i) = \ell_i(\parm)$ is the likelihood contribution by
some observation $\rY_i$. By definition, the true unknown parameter $\parm(\rx)$
has minimal expected risk and thus maximises the expected log-likelihood, \ie
\begin{eqnarray*}
\parm(\rx) = \argmax_{\parm \in \Theta} \ell(\parm \mid \rX = \rx).
\end{eqnarray*}
Our random forest-type estimator of the expected conditional log-likelihood given $\rx$ for
a fixed set of parameters $\parm$ is now
\begin{eqnarray*}
\hat{\ell}_N(\parm \mid \rX = \rx) = \frac{\sum_{i = 1}^N w_{\text{Forest}, i}^N(\rx) \ell(\parm,
\rY_i)}{\sum_{i = 1}^N w_{\text{Forest}, i}^N(\rx)}.
\end{eqnarray*}
Under the respective conditions  on the distribution of $\rX$ and the joint
distribution of $\rY, \rX$ given by \cite{Lin_Jeon_2006},
\cite{Biau_Devroye_2010}, or \cite{Biau_2012}, this estimator is consistent
for all $\parm \in \Theta$ 
\begin{eqnarray*}
\Ex_{\rY \mid \rX = \rx}\left( \ell(\parm \mid \rX = \rx) -
\hat{\ell}_N(\parm \mid \rX = \rx) \right)^2 \rightarrow 0
\end{eqnarray*}
(the result being derived for non-adaptive random forests). This result
gives us consistency of the conditional log-likelihood function 
\begin{eqnarray*}
\hat{\ell}_N(\parm \mid \rX = \rx) \stackrel{\Prob}{\rightarrow}
\ell(\parm \mid \rX = \rx) \quad \forall \parm \in \Theta.
\end{eqnarray*}
The forest conditional parameter function
$\hat{\parm}_\text{Forest}^N(\rx)$ is consistent when
\begin{eqnarray*}
\Prob_\parm( \hat{\ell}_N(\parm_1 \mid \rX = \rx) < \hat{\ell}_N(\parm \mid \rX = \rx)) \stackrel{\Prob}{\rightarrow} 1
\end{eqnarray*}
as $N \rightarrow \infty$ for all $\parm_1$ in a neighbourhood of
$\parm$. The result 
$\hat{\parm}_\text{Forest}^N(\rx) \stackrel{\Prob}{\rightarrow} \parm(\rx)$
can be shown under the assumptions
regarding $\ell$ given by \cite{Hothorn_Moest_Buehlmann_2016}, especially continuity 
in $\parm$. Because the conditional log-likelihood $\hat{\ell}_N(\parm \mid
\rX = \rx)$ is a conditional
mean-type estimator of a transformed target $\rY$, future theoretical
developments in the asymptotic analysis of more realistic random forest-type
algorithms based on nearest neighbour weights will directly carry over to transformation forests.

It is worth noting that some authors studied properties of random forests in
regression models of the form $\rY = \reg(\rx) + \varepsilon$ where the
conditional variance $\V(\rY \mid \rX = \rx)$ does not depend on $\rx$.
This is in line with the ANOVA split criterion implemented in Breiman and
Cutler's random forests \citep{Breiman_2001}. The split procedure applied in transformation trees
is, as will be illustrated in the next section, able to detect changes in
higher moments. Thus, transformation forests might be a way to relax the
assumption of additivity of signal and noise in the future.

\section{Empirical Evaluation} \label{sec:empeval}


\NEW{Transformation forests were evaluated empirically, comparing this
novel member of the random forest family to established procedures using
artificial data generating processes. The data scenarios controlled the
variation of several properties of interest: type of conditional parameter
function, types of effect, and model complexity in low and high dimensions.
The corresponding hypotheses to be assessed are:
\begin{description}
\item[H1:] \textbf{Type of Conditional Parameter Regression.}
\begin{description}
  \item[H1a:] \textbf{Tree-Structured Conditional Parameter Function.} Transformation trees and forests are able to identify subgroups
          associated with different transformation models, \ie subgroups formed by a recursive
	  partition (or tree) in predictor variables $\rx$ corresponding to different parameters
	  and thus different conditional distributions $\ProbYx$.
  \item[H1b:] \textbf{Non-Linear Conditional Parameter Function.} Transformation forests are able to 
              identify conditional distributions $\ProbYx$ whose parameters depend 
              on predictor variables $\rx$ in a smooth non-linear way.
\end{description}
\item[H2:] \textbf{Type of Effect.}
\begin{description}
  \item[H2a:] \textbf{No Effect.} In a non-informative scenario with $\ProbYx = \Prob_\rY$
           (i.e., mean and all higher moments constant) transformation trees perform as good as the
           unconditional maximum likelihood estimator. Thus, there is no (pronounced) overfitting.
  \item[H2b:] \textbf{Location Only.} Transformation trees and forests perform as good as classical regression
           trees and forests when higher moments of the conditional distribution 
          $\ProbYx$ are constant. 
  \item[H2c:] \textbf{Unlinked Location and Scale.} Transformation trees and forests outperform classical regression
           trees and forests when higher moments of the conditional distribution 
          $\ProbYx$ are varying in a way that is \emph{not} linked to variations in the mean.
  \item[H2d:] \textbf{Linked Location and Scale.} Transformation trees and forests perform as good as classical regression
           trees and forests when higher moments of the conditional distribution 
          $\ProbYx$ are varying but in a way that is linked to the mean.
\end{description}
\item[H3:] \textbf{Model Complexity.}
  \begin{description}
  \item[$P = 2$:] Transformation trees and forests with linear transformation function
           $\h$, \ie with $P = 2$ parameters, perform best for conditionally normal target variables. Transformation trees and forests
           with non-linear transformation function $\h$ perform slightly worse in
           this situation.
  \item[$P = 6$:] Transformation trees and forests with non-linear transformation function $\h$, \ie with $P = 6$ parameters of a Bernstein polynomial of order five,
           outperform transformation trees and forests with linear
           transformation function for conditionally non-normal target variables.
  \end{description}
\item[H4:] \textbf{Dimensionality.} Transformation forests stabilise
           transformation trees in the presence of high-dimensional non-informative
           predictor variables.
\end{description}}

\subsection{Data Generating Processes}

\NEW{Two data generating processes corresponding to \textbf{H1a} and
\textbf{H1b} were studied.  
The first problem implements simple binary splits in the conditional mean
and/or
conditional variance of a normal target allowing a direct comparison of the split criteria employed by
classical and transformation trees.  The second problem is inspired by the
``Friedman~1'' benchmark problem \citep{Friedman_1991}, and implements smooth
non-linear conditional mean and variance functions for normal targets,
in order to provide a more complex and more realistic scenario.}

\paragraph{Tree-Structured Conditional Parameter Function (H1a)}

\NEW{The conditional normal target
\begin{eqnarray} \label{sim:2d}
\rY \mid \rX = \rx \sim \ND\left(\mu_\text{Tree}(\rx), \sigma_\text{Tree}(\rx)^2\right)
\end{eqnarray}
depends on tree-structured conditional mean and variance functions $\mu(\rx)$ and
$\sigma(\rx)^2$ according to 
four different setups (corresponding to hypotheses \textbf{H2a--c}):
\begin{center}
\begin{tabular}{ccc} \hline
  & $\mu_\text{Tree}(\rx)$ & $\sigma_\text{Tree}(\rx)$ \\ \hline \hline
\textbf{H2a} & $0$ & $1$ \\
\textbf{H2b} & $I(x_1 > .5)$ & $1$ \\
\textbf{H2c} & $0$ & $(1 + I(x_2 > .5))$ \\
\textbf{H2c} & $I(x_1 > .5)$ & $(1 + I(x_2 > .5))$ \\ \hline
\end{tabular}
\end{center}
All predictors $\rX = (X_1, \dots, X_7)$ are independently uniform on $[0, 1]$
in the low-dimensional case (two informative and five noise variables) 
and $\rX = (X_1, \dots, X_{52})$ in the high-dimensional case (two
informative and $50$ noise variables, \textbf{H4}).}

\NEW{For the evaluation of hypothesis \textbf{H2d} 
we studied the same setup as above but for conditionally log-normal targets with
\begin{eqnarray} \label{sim:ln-2d}
\rY^\prime = \exp(\rY).
\end{eqnarray}
Here, the conditional mean of the target variable $\rY^\prime$ depends both
on the underlying conditional mean $\mu(\rx)$ of $\rY$ and the corresponding
conditional variance $\sigma(\rx)^2$:
\begin{eqnarray*}
\Ex(\rY^\prime \mid \rX = \rx) = \exp(\mu(\rx) + \sigma(\rx)^2/2)
\end{eqnarray*}
It is important to note that the true transformation function $\h$ in model
(\ref{sim:ln-2d}) is a scaled and shifted log-transformation.  Unlike the
true linear transformation function $\h$ in model (\ref{sim:2d}), which can
be exactly fitted by the linear and Bernstein parameterisations of the
transformation function in transformation trees and forests,
the true log-transformation cannot be approximated well by the basis
functions $\basisy$. Therefore,
no competitor in this simulation experiment is able to exactly recover the
true data generating process.}

\paragraph{Non-Linear Conditional Parameter Function (H1b)}

\NEW{The data generating process
\begin{eqnarray} \label{sim:fm}
\rY \mid \rX = \rx \sim \ND\left(\mu_\text{Nonlin}(\rx), \sigma_\text{Nonlin}(\rx)^2\right)
\end{eqnarray}
with all predictors $\rX = (X_1, \dots, X_{15})$ from independent uniform
distributions on $[0, 1]$ in the low-dimensional case (ten informative
and five noise variables) and $\rX = (X_1, \dots, X_{60})$ in the
high-dimensional case (ten informative and $50$ noise variables, \textbf{H4}) is inspired
by the ``Friedman~1'' benchmarking problem \citep{Friedman_1991}. 
This original benchmark problem is conditional normal with a 
conditional mean function depending on five uniform predictor variables
\begin{eqnarray*}
\text{Friedman1}(x_1, x_2, x_3, x_4, x_5) = 10 \sin(\pi x_1 x_2) + 20 (x_3 - .5)^2 + 10 x_4 + 5 x_5
\end{eqnarray*}
and constant variance. For our experiments, we scaled the output of $\text{Friedman1}$ 
to the $[-1.5, 1.5]$ interval and denote this scaled function as $\text{Friedman1}^\star$.
Model (\ref{sim:fm}) is conditionally normal with potentially non-constant
conditional mean function 
\begin{eqnarray*}
\mu_\text{Nonlin}(\rx) = \text{Friedman1}^\star(x_1, x_2, x_3, x_4, x_5)
\end{eqnarray*}
and potentially non-constant conditional variance function 
\begin{eqnarray*}
\sigma_\text{Nonlin}(\rx) = \exp(\text{Friedman1}^\star(x_6, x_7, x_8, x_9,
x_{10}))^2.
\end{eqnarray*}
The latter function is based on
an additional set of five uniformly distributed predictor variables and thus
the conditional mean and variance function are not linked (\textbf{H2c}).}

Again, we considered all setups corresponding to \textbf{H2a--c} and \textbf{H4}, including 
the non-informative case with constant mean and variance:
\begin{center}
\begin{tabular}{ccc} \hline
  & $\mu_\text{Nonlin}(\rx)$ & $\sigma_\text{Nonlin}(\rx)$ \\ \hline \hline
\textbf{H2a} & $0$ & $1$ \\
\textbf{H2b} & $\text{Friedman1}^\star(x_1, x_2, x_3, x_4, x_5)$ & $1$ \\
\textbf{H2c} & $0$ & $\exp(\text{Friedman1}^\star(x_6, x_7, x_8, x_9, x_{10}))$ \\
\textbf{H2c} & $\text{Friedman1}^\star(x_1, x_2, x_3, x_4, x_5)$ & $\exp(\text{Friedman1}^\star(x_6, x_7, x_8, x_9, x_{10}))$ \\ \hline
\end{tabular}
\end{center}

\NEW{Hypothesis \textbf{H2d} for non-linear conditional parameter 
functions (\textbf{H1b}) was studied in the log-normal model
\begin{eqnarray} \label{sim:ln-fm}
\rY^\prime = \exp(\rY).
\end{eqnarray}
and the remarks to model (\ref{sim:ln-2d}) stated above also apply here.}

\subsection{Competitors}

\NEW{For testing the hypotheses \textbf{H1--H4}, we compared the performance of transformation
trees and forests with linear and non-linear transformation functions $\h$
to the performance of 
conditional inference trees \citep{Hothorn_Hornik_Zeileis_2006} and
conditional inference forests \citep{Strobl_Boulesteix_Zeileis_2007} as representatives of unbiased
recursive partitioning and to Breiman and Cutler's random forests \citep{Breiman_2001} as an 
representative of exhaustive search procedures. In more detail, we compared the
performance of the following methods:
\begin{description}
\item[CTree:] Conditional inference trees with internal stopping by default parameters.
\item[TTree:] Transformation trees, either with linear ($P = 2$ parameters) or
              non-linear ($P = 6$ parameters of a Bernstein polynomial) transformation functions.
	      Tree-growing parameters are identical to those from CTree.
\item[CForest:] Conditional inference forests with \texttt{mtry} equal to one third of the
              number of predictor variables. Trees were grown without internal stopping
              until sample size constraints were met.
\item[RForest:] Breiman and Cutler's random forests with tree-growing parameters analogous to CForest
              (i.e., same \texttt{mtry} and stopping based on sample size constraints).
\item[TForest:] Transformation forests, either with linear ($P = 2$) or non-linear ($P = 6$)
              transformation functions, and tree-growing parameters analogous to CForest and RForest.
\end{description}}
\NEW{
See Table~\ref{tab:comp} for a schematic overview of all competitors and
Appendix~``Computational Details'' for the exact tree-growing parameter specifications.

In order to allow a fair comparison on the same scale, trees and
forests obtained from the classical methods, \ie conditional inference trees
and forests and Breiman and Cutler's random forests, were used to estimate
conditional parameter functions (\ref{fm:treeweights}) and
(\ref{fm:RFweights}) in the same way as for transformation trees and
forests: We first fitted trees and forests using the reference
implementations of the corresponding methods and, second, computed the
corresponding conditional weight functions, which allowed estimation of
conditional parameter functions $\parm^N_\text{CTree},
\parm^N_\text{CForest}$, and $\parm^N_\text{RForest}$ in the third step.  It
should be noted that the combination of Breiman and Cutler's random forests
with transformation models in our RForest variant is conceptually very
similar to quantile regression forests. 
\cite{Meinshausen_2006,pkg:quantregForest} uses Breiman and Cutler's random
forest to build the trees.  The only difference to our RForest variant is
that aggregation in quantile regression forests takes place via the weighted
empirical conditional cumulative distribution function with weights
$w^N_\text{randomForest}$, see Formula~(\ref{fm:QRF}), instead of the
application of a smooth conditional distribution function corresponding to a
transformation model.}

\begin{table}
\begin{center}
\begin{tabular}{p{2cm}p{2cm}p{2.75cm}|p{3.5cm}p{3.5cm}} \hline \noalign{\smallskip}
\multicolumn{3}{c|}{Tree Growing} & \multicolumn{2}{c}{Model Complexity} \\
Variable & Split & Split & Linear ($P = 2$) & Non-Linear ($P = 6$) \\
Selection & Search & Criterion &  &  \\ \hline \hline \noalign{\smallskip}
exhaustive & exhaustive RSS & MSE & RForest ($P = 2$) & RForest ($P = 6$) \\ \hline \noalign{\smallskip}
unbiased (inference based) & maximally selected score test & location &
\parbox[t]{3cm}{CTree ($P = 2$) \\ CForest ($P = 2$)} & \parbox[t]{3cm}{CTree ($P = 6$) \\ CForest ($P = 6$)} \\
 &  & location/scale & \parbox[t]{3cm}{TTree ($P = 2$) \\ TForest ($P = 2$)} &  \\
 &  & \parbox[t]{2.75cm}{higher moments \\ (Bernstein)} &  & \parbox[t]{3cm}{TTree ($P = 6$) \\ TForest ($P = 6$)} \\ \cline{2-5} \noalign{\smallskip}
 & exhaustive likelihood & location/scale & TTree ($P = 2$, exh) &  \\
 &  & \parbox[t]{2.75cm}{higher moments \\ (Bernstein)} &  & TTree ($P = 6$, exh) \\ \hline
\end{tabular}
\end{center}
\caption{Competitor Overview. All competitors ordered with respect to
variable and split selection procedures as well as complexity of the
underlying transformation function $\h$. Abbreviations: Mean-squared error
(MSE), residual sum of squares (RSS), exhaustive (exh). \label{tab:comp}}
\end{table}

\subsection{Performance Measures}

\NEW{
The primary performance measure is the out-of-sample log-likelihood
because it assesses the whole predicted distribution in a ``proper''
way \citep{Gneiting_Raftery_2007}. To adjust for sampling variation,
the log-likelihood of the true data generating process is employed
as the reference measure. More precisely, the negative log-likelihood
difference, that is the negative log-likelihood of a competitor minus
the negative log-likelihood of the true data generating process, was
evaluated for the $N = 250$ observations of the validation sample.
Conditional medians and prediction intervals are of
additional interest and we also compared their performance by the out-of-sample
check risk corresponding to the $10\%$, $50\%$ (absolute error) and $90\%$
quantiles in reference to the true data generating process.
A direct comparison of coverage and lengths of prediction
intervals is not considered as it would only be valid or useful for
a given configuration of the predictor variables. This is termed
``conditional coverage'' vs.\ ``sample coverage'' in \cite{Mayr_Hothorn_Fenske_2012}
or considered as maximising forecast sharpness only subject to
calibration in the proper scoring rules literature \citep{Gneiting_Balabdaoui_Raftery_2007}.
}

\subsection{Results: Tree-Structured Conditional Parameter Function (H1a)}

\begin{figure}
\begin{center}
\includegraphics{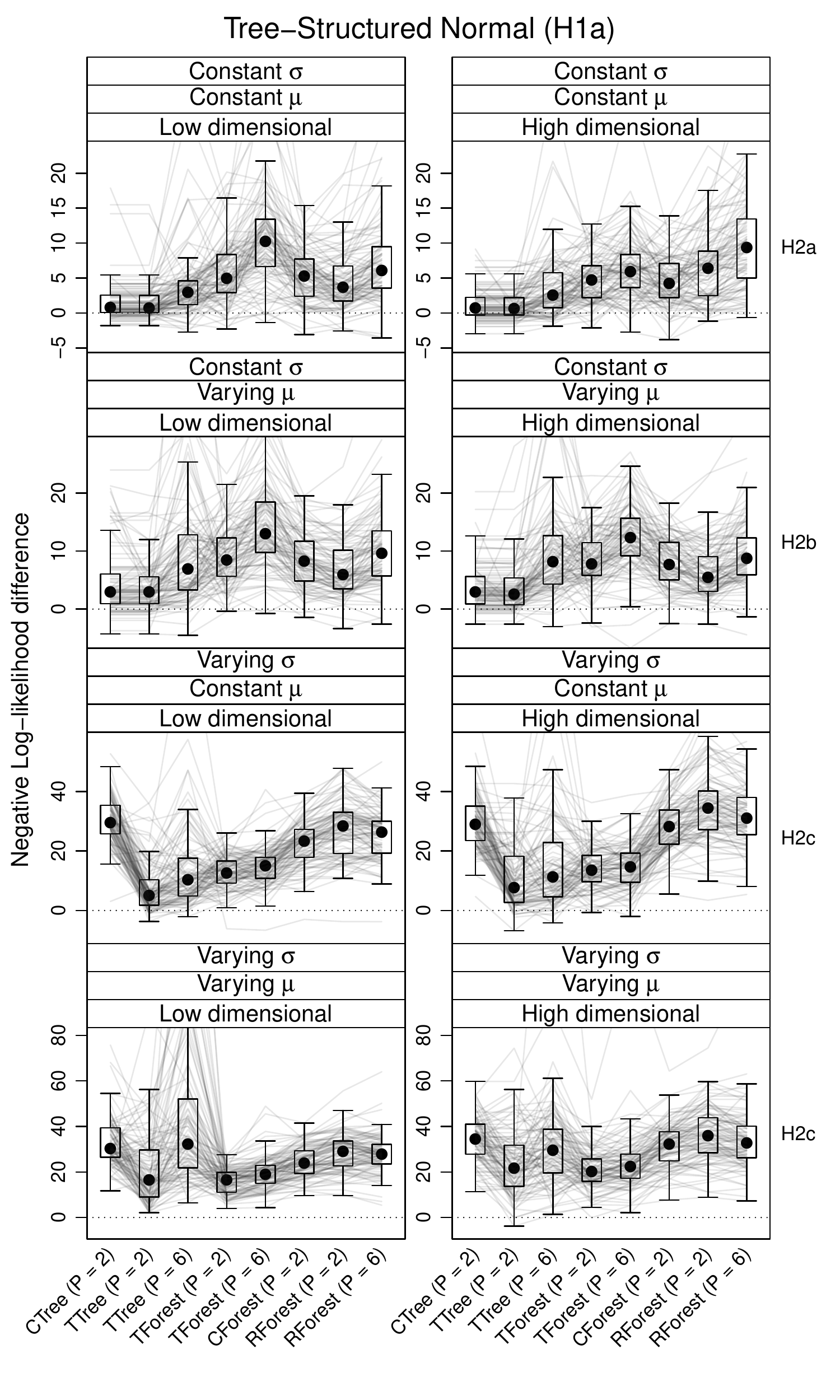}
\caption{Simulation Model (\ref{sim:2d}).  Negative log-likelihood
differences for trees
and forests in a conditional normal model with potential jumps in mean and
variance.  The negative log-likelihood
difference was computed as the out-of-sample
negative log-likelihood of each competitor minus the negative log-likelihood of the true
data generating process. Outliers were not plotted.
\label{2d-ll}}
\end{center}
\end{figure}

\NEW{
Given the type of conditional parameter function (here: tree, \textbf{H1a}) all other properties
of the data generating process are varied and assessed, summarising the results with
parallel coordinate displays and superimposed boxplots of the negative log-likelihood
differences
(see Figure~\ref{2d-ll}). These were obtained from $100$ pairs of learning samples
(size $N = 250$) and validation samples, using a normal dependent variable in the first step.
This allows to assess the type of effect
(mean and/or higher moments) in the rows of the panels (\textbf{H2a--c}), the dimensionality
(\textbf{H4}) in the columns of the panels, and the complexity (\textbf{H3}, $P = 2$ vs.~$6$) along the
$x$-axes.

In the situation where all predictor variables were non-informative (\textbf{H2a}, top row of
Figure~\ref{2d-ll}), CTree ($P = 2$) and TTree ($P = 2$) were most resistant to
overfitting; this effect is due to the test-based internal stopping
of the unbiased tree methods compared here. TTree ($P = 6$) with
non-linear transformation function had slightly larger negative log-likelihood
differences
due to the increased model complexity (\textbf{H3}). Moreover, if model complexity
is further increased by considering forests instead of trees, all random forest
variants exhibit some more pronounced overfitting behaviour.

Under the simple change in the mean (\textbf{H2b}, second row in Figure~\ref{2d-ll}), 
CTree ($P = 2$) and TTree ($P = 2$) were able to detect this split best. TTree ($P = 6$)
and all random forest variants performed less well in this situation.
A variance change (\textbf{H2c}, third row in Figure~\ref{2d-ll}) lead to smallest
negative log-likelihood difference
and thus superior performance for all transformation
trees and forests as compared to the trees and forests splitting only based on the mean.
TTree ($P = 2$) performed best while none of the classical procedures
seemed to be able to properly pick up this variance signal. The aggregation of
multiple transformation trees lead to decreased performance, this effect was
also visible in Figure~\ref{fig:exIntro} (which was based on the same data
generating process (\ref{sim:2d})).

When changes in both mean and variance were present (\textbf{H2c}, fourth row in
Figure~\ref{2d-ll}), transformation forests with linear transformation
function TForest ($P = 2$) performed as good as the corresponding TTree in the
low-dimensional setup but better than all other procedures in the high-dimensional
setup with $50$ non-informative variables (\textbf{H4}). This effect might be due to a too
restrictive inference-based early stopping in TTree. TTree ($P = 6$) showed some
extreme outliers (\textbf{H3}, visible in the parallel coordinates in Figure~\ref{2d-ll})
which were due to convergence problems. The corresponding transformation forests TForest
($P = 6$), however, did not experience such problems and thus seemed to stabilise the
trees.}

\NEW{In summary, the results with respect to our hypotheses were:
\begin{description}
  \item[H1a:] Transformation trees reliably recover tree-structured conditional
    parameter functions in both mean and variance.
  \item[H2a:] Transformation trees are rather robust to overfitting when there is no effect
    while transformation forests (like all other random forests) exhibit some overfitting.
  \item[H2b:] Transformation trees and forests perform comparably to their
    classical counterparts.
  \item[H2c:] Transformation trees and forests outperform their classical counterparts
    if there are only variance effects or variance effects that are not linked to the mean.
  \item[H3:] For normal responses transformation trees and forests with linear transformation
    function ($P = 2$) consistently perform better than the more complex Bernstein polynomials
    ($P = 6$).
  \item[H4:] Transformation forests stabilise the transformation trees in high-dimensional
    settings.
\end{description}}

\begin{figure}
\begin{center}
\includegraphics{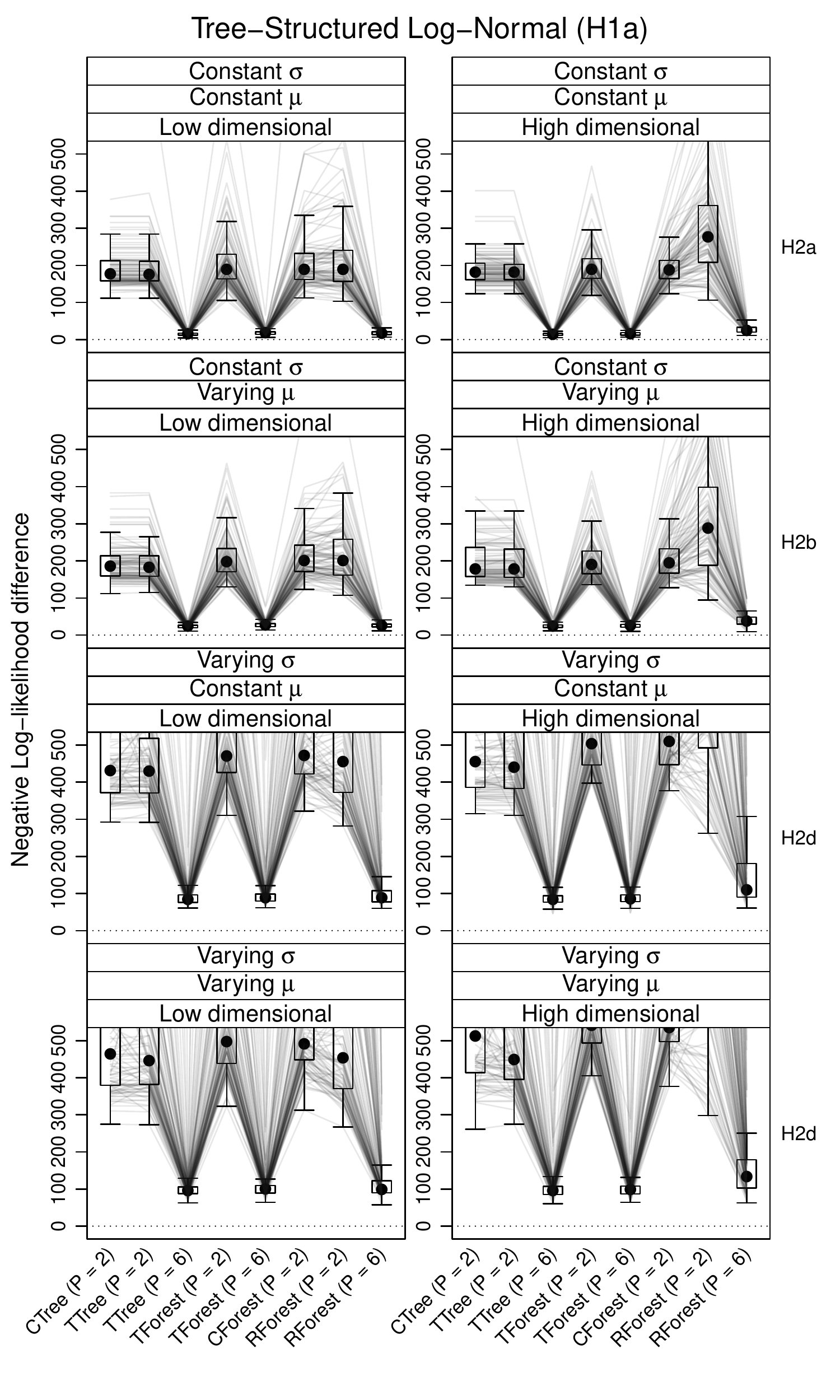}
\caption{Simulation Model (\ref{sim:ln-2d}).  Negative log-likelihood
differences for trees
and forests in a conditional log-normal model with potential jumps in mean
and variance.  The negative log-likelihood difference was computed as the out-of-sample
negative log-likelihood of each competitor minus by the negative log-likelihood of the true
data generating process. Values larger than two were not plotted.
\label{ln-2d-ll}}
\end{center}
\end{figure}

\NEW{As a next step, the same simulation experiments were considered using a log-normal
target variable instead of the normal variable employed above. Figure~\ref{ln-2d-ll}
depicts the negative log-likelihood differences for this setup, based on $100$
learning samples of size $N = 2500$. Using this highly skewed
distribution affects the results regarding the following two hypotheses:
\begin{description}
  \item[H3] All models with complexity $P = 2$ are clearly not appropriate anymore
    as they cannot capture the skewness. Consequently, all models based on the more
    flexible Bernstein polynomials with $P = 6$ outperform all other methods.
  \item[H2d] The classic RForest ($P = 6$), \ie the combination of Breiman and Cutler's random
    forests with a subsequent flexible transformation model, performs almost on par with
    transformation trees and forests even when there are changes in the variance only.
    The reason is that any changes in the variance are always also linked to
    changes in the mean due to the skewness of the distribution.
\end{description}}

\NEW{Qualitatively the same conclusions can be drawn when assessing the competing
methods based on predictions of the conditional $10\%$ quantiles (Figure~\ref{2d-L1}
and~\ref{ln-2d-L1} for normal and log-normal targets, respectively),
$50\%$ quantiles (Figure~\ref{2d-L5} and~\ref{ln-2d-L5}), and $90\%$ quantiles
(Figure~\ref{2d-L9} and~\ref{ln-2d-L9}). However, the differences are less pronounced
for the $50\%$ quantiles (medians, corresponding to the absolute errors). Note also that
combining predictions of $10\%$ and $90\%$ quantiles amounts to $80\%$ prediction intervals.}

\NEW{By and large, all empirical results in this section conformed with our
hypotheses \textbf{H1--4}, suggesting a stable behaviour of transformation trees and
forests, especially with appropriate linear transformation function for
normal targets, in these very simple situations. The next section proceeds to a
less idealised scenario with non-linear conditional parameter functions defining mean
and/or variance.}

\subsection{Results: Non-Linear Conditional Parameter Function (H1b)}

\begin{figure}
\begin{center}
\includegraphics{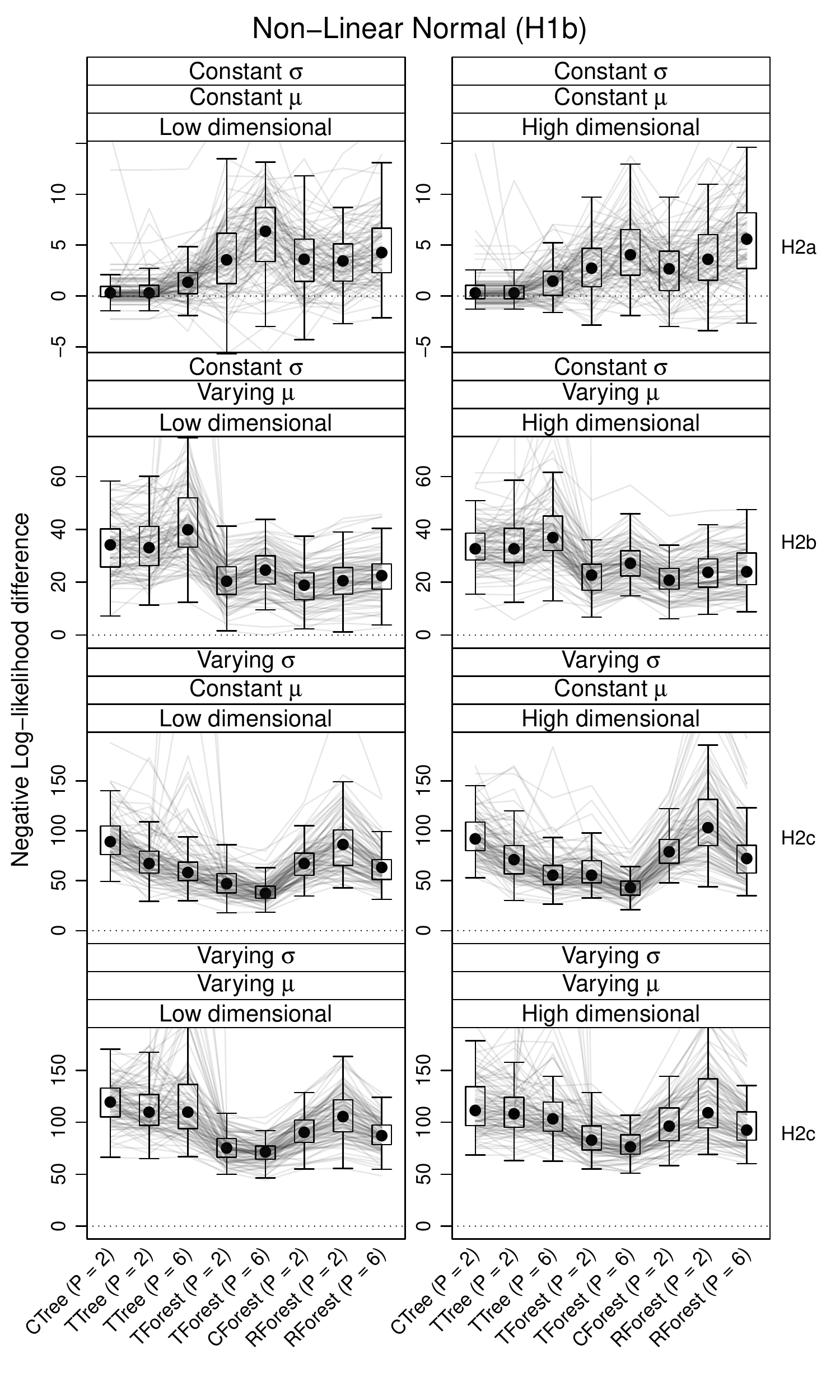}
\caption{Simulation Model (\ref{sim:fm}).  Negative log-likelihood differences for trees
and forests in a conditional normal model with non-linear functions defining
mean and variance.  The negative log-likelihood difference was computed as the
out-of-sample negative log-likelihood of each competitor minus the
negative log-likelihood of the true data generating process. Outliers were not
plotted.
\label{fm-ll}}
\end{center}
\end{figure}

\NEW{The same hypotheses were assessed as in the previous section
but for non-linear Friedman1-type conditional parameter functions instead of the
tree-structured functions considered previously. More specifically, Figures~\ref{fm-ll}
and~\ref{sim:ln-fm} depict the negative log-likelihood differences based on $100$
learning samples with normally-distributed targets ($N = 500$) and log-normally-distributed
targets ($N = 2500$), respectively. We summarise the results as follows.
\begin{description}
  \item[H1b:] When a signal was present (rows~2--4), all random forest variants outperformed single trees
    under normality. Under non-normality, this still holds for the random forest variants
    combined with flexible models ($P = 6$).
  \item[H2a:] When there is no effect (top rows), CTree ($P = 2$) and TTree ($P = 2$) showed
    best resistance to overfitting under normality. Under non-normality, TTree ($P = 6$)
    still shows this behavior but the corresponding forests also perform similarly well.
  \item[H2b:] All forest variants performed similarly well when predictor variables only had an
    effect on the mean (second rows).
  \item[H2c:] Under normality, transformation forests performed best when some of the
    predictor variables also affected the variance (rows~3--4), where the classical procedures
    were not able to capture these changes appropriately.
  \item[H2d:] Under non-normality, transformation forests (with $P = 6$) still perform
    best (rows~3--4). However, the classical RForest also perfoms well albeit with a much
    larger variance than TForest.
  \item[H3:] Under non-normality, all trees and forests combined with flexible Bernstein
    polynomials ($P = 6$) clearly outperform all other methods. Under normality,
    the flexible models with $P = 6$ were sometimes slightly worse than the $P = 2$
    models but often also a little bit better.
  \item[H4:] In many situations the picture in low-dimensional settings (left column)
    is quite similar to that in high-dimensional scenarios (right column). However,
    sometimes it can be seen that transformation forests stabilise transformation trees
    in the presence of high-dimensional non-informative predictor variables.
\end{description}
}

\begin{figure}
\begin{center}
\includegraphics{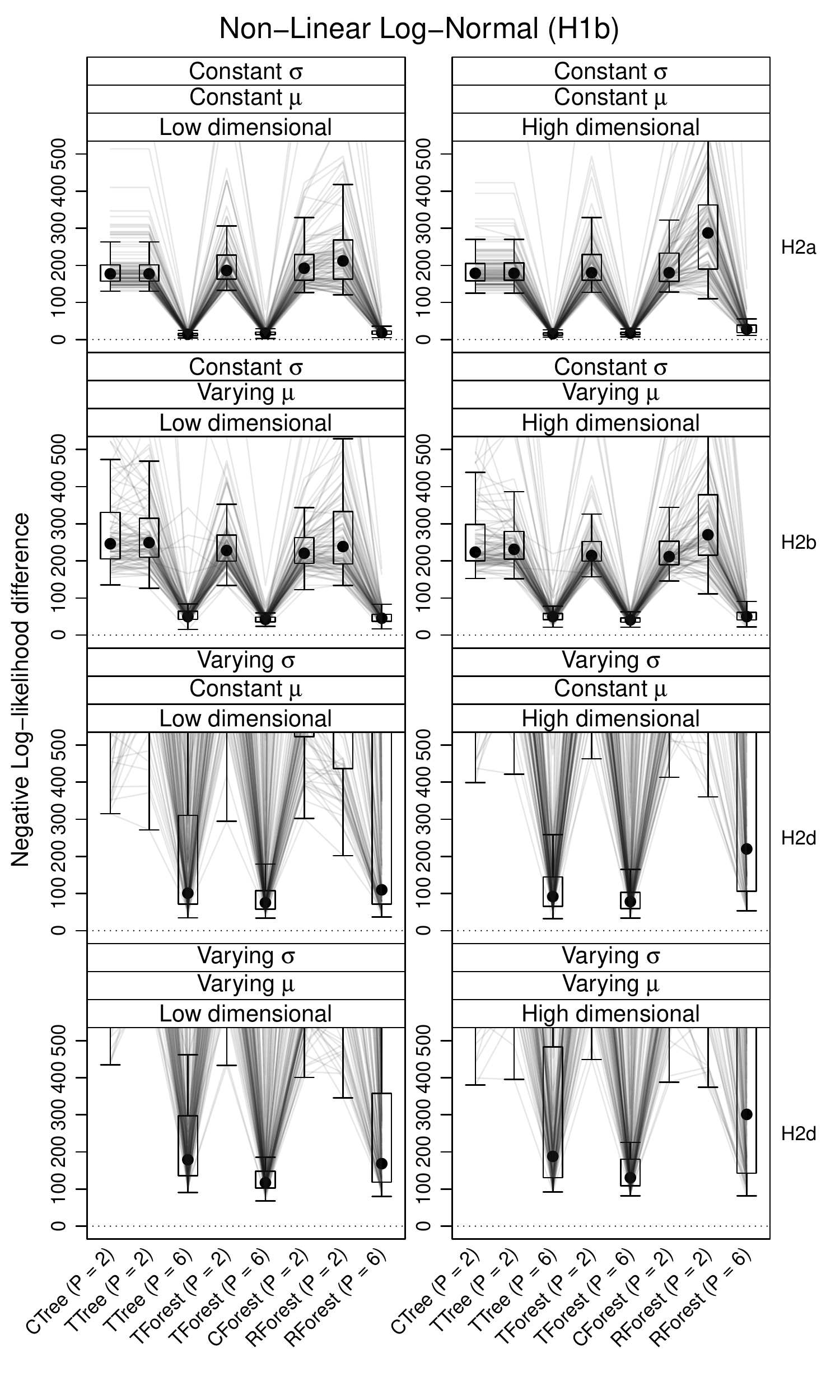}
\caption{Simulation Model (\ref{sim:ln-fm}).  Negative log-likelihood differences for
trees and forests in a conditional log-normal model with non-linear
functions defining mean and variance.  The negative log-likelihood
difference was computed
as the out-of-sample negative log-likelihood of each competitor minus the
negative log-likelihood of the true data generating process. Values larger than two
were not plotted.
\label{ln-fm-ll}}
\end{center}
\end{figure}

\NEW{As before, qualitatively the same patterns could be observed for the corresponding
$10\%$, $50\%$, and $90\%$ check risks (Figures~\ref{fm-L1}--\ref{fm-L9} and
Figures~\ref{ln-fm-L1}--\ref{ln-fm-L9}, respectively) and thus prediction intervals.
In summary, our hypotheses \textbf{H1--4} were found to describe the
behaviour of transformation trees and forests in this more complex setup well. The
loss of using an overly complex model, such as a transformation model with $P = 6$,
was tolerable in the simple normal setups but the gains, especially when parameters of
a skewed target depend on the predictor variables, was found to be quite substantial.}

\subsection{Illustration: Swiss Body Mass Indices}


\NEW{Finally, to conclude this section, we illustrate the applicability of transformation trees and forests in a 
realistic situation by modelling the conditional body mass index (BMI = weight
(in kg) / height (in m)$^2$) distribution for Switzerland, based on
$16{,}427$ individuals aged between $18$ and $74$ years from the 2012 Swiss
Health Survey \citep{SHS_2012}.  The predictor variables included smoking,
sex, age, and a number of ``lifestyle variables'' $\rx$: fruit and vegetable
consumption, physical activity, alcohol intake, level of education,
nationality and place of residence.  Smoking status was categorised into
never, former, light, moderate, and heavy smokers.  A more detailed
description of this data set can be found in \cite{Lohse_Rohrmann_Faeh_2017}
and extended transformation models for body mass indices are discussed 
by \cite{Hothorn_2018}.}

\NEW{The conditional transformation model underlying transformation trees and
transformation forests
\begin{eqnarray*}
\Prob(\text{BMI} \le \ry \mid \text{sex}, \text{smoking}, \text{age}, \rx) = 
    \pZ\left(\bern{5}(\ry)^\top \parm(\text{sex}, \text{smoking}, \text{age}, \rx)\right),
\end{eqnarray*}
assumes that each conditional distribution is parameterised in terms of a
Bernstein polynomial with $P = 5$.  The parameters $\parm$ of
this polynomial, however, might depend on the predictor variables in a
potentially complex way, featuring interactions and non-linearities. 
Transformation trees and forest allow such conditional parameter functions
$\parm$, and thus the corresponding conditional BMI distributions, to be
estimated in a black-box manner without the necessity to \textit{a priori}
specify any structure of $\parm$ \citep[models assuming such structures are
discussed in][]{Hothorn_2018}.}

\begin{sidewaysfigure}
\includegraphics{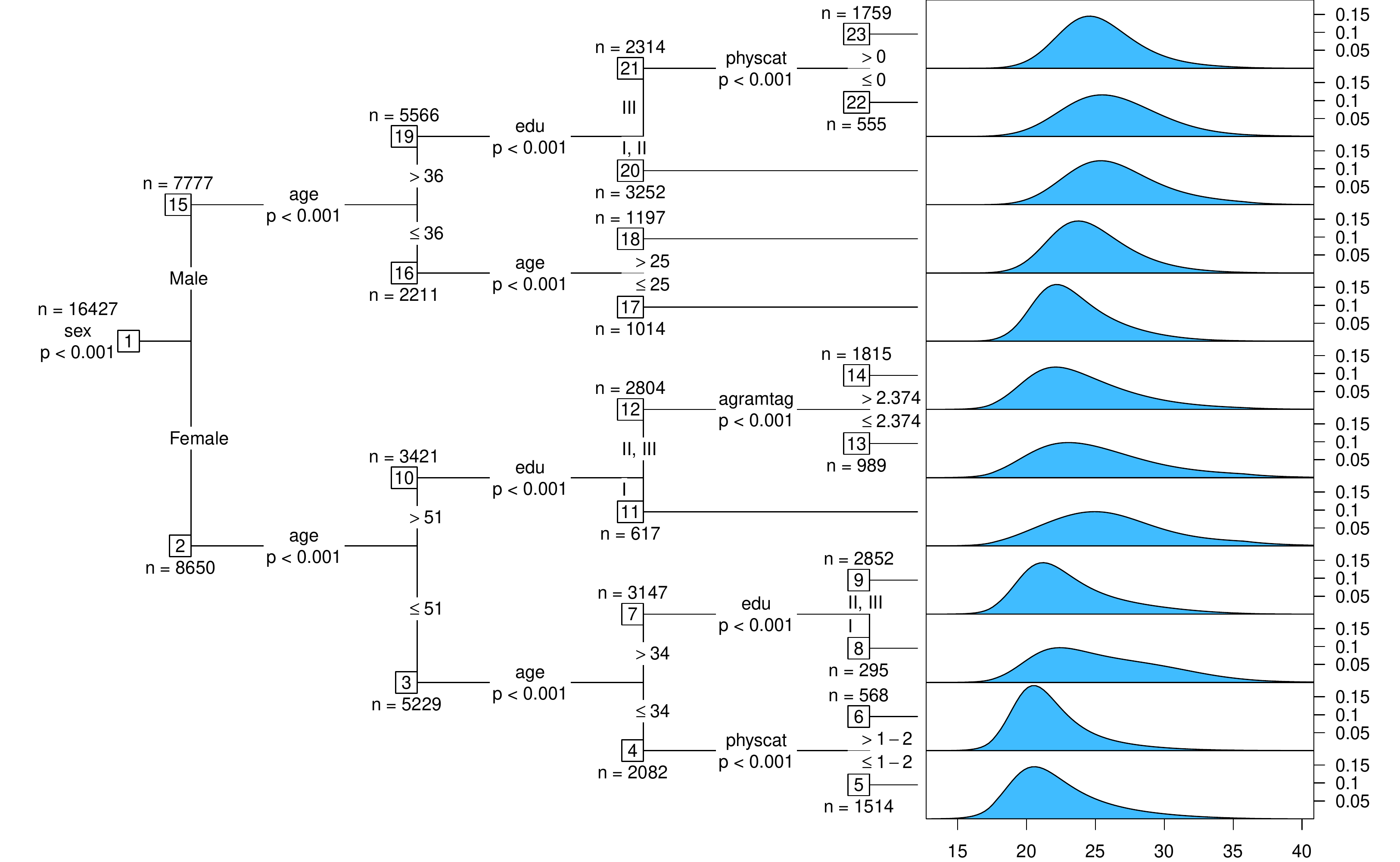}
\caption{Body Mass Index (BMI). The conditional BMI distributions (depicted in
         terms of their densities) are given in each subgroup of the
         transformation tree (featuring a non-linear transformation function) corresponding
         to the terminal nodes of the tree. Variables: education (edu) at
         levels mandatory (I), secondary (II) and tertiary (III); 
         alcohol intake (agramtag). \label{BMI-tree-plot}}
\end{sidewaysfigure}

\NEW{The in-sample negative log-likelihood of the tree presented in Figure~\ref{BMI-tree-plot} is
$43079.42$.  The first split was in sex, so in fact two
sex-specific models are given here.  Four age groups ($\le 34$, $(34, 51]$,
$> 51$) for females and three age groups ($\le 25$, $(25, 36]$,
$> 36$) for males were distinguished.  Education contributed to understanding
the BMI distribution of females and males.  Location, scale and shape of the
conditional BMI distributions varied considerably. Higher BMI variability was
linked to higher average BMI values. Mean and variance increased
with age, and higher-educated people tended to have lower BMI values.  These
are interesting insights, but this tree model is, of course, very rough.}

\begin{figure}
\begin{center}
\includegraphics{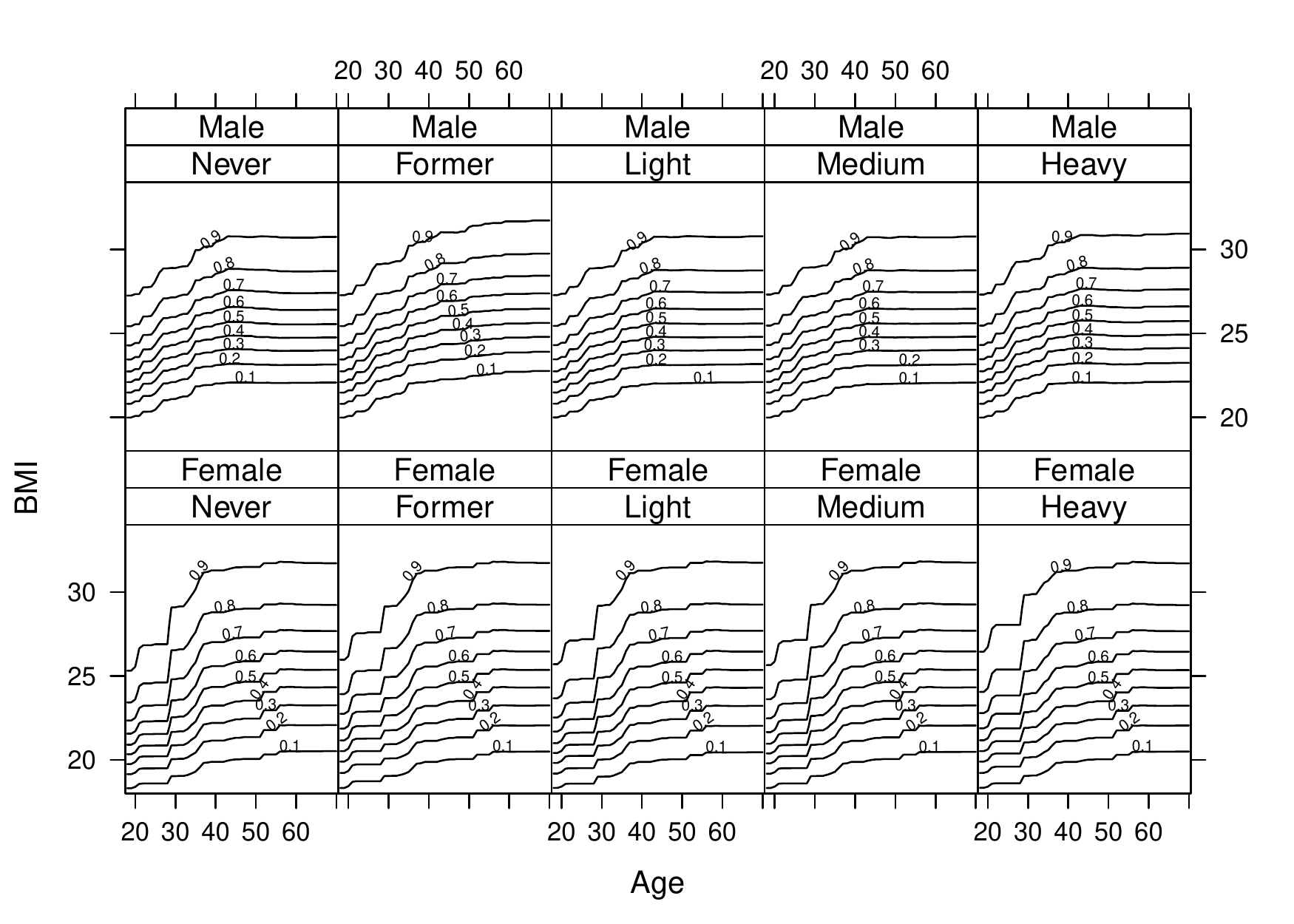}
\caption{Body Mass Index (BMI). Partial dependency plot of conditional deciles estimated
         by a transformation forest with non-linear transformation function. 
         \label{BMI-forest-plot}}
\end{center}
\end{figure}

\NEW{A transformation forest allows less rough conditional parameter functions
$\parm$ to be estimated.  The negative log-likelihood was
$42520.18$ and thus a substantial improvement over
the negative log-likelihood $43079.42$ of the
transformation tree.}

\NEW{However, such black-box models are rather difficult to understand in terms of
the impact of the predictor variables on the conditional BMI distribution. 
We used a partial dependency plot for conditional deciles to visualise the
association between sex, smoking, age and BMI as estimated by the
transformation forest (Figure~\ref{BMI-forest-plot}).  In general, the median
BMI increases with age, as does the BMI variance.  For males, there seemed
to be a level-effect whose onset depends on smoking category.  Females
tended to higher BMI values, and the variance was larger compared to males. 
There seemed to be a bump in BMI values for females, roughly around $30$
years.  This corresponds to mothers giving birth to their first child around
this age.  It is important to note that the right-skewness of the
conditional BMI distributions renders conditional normal distributions
inappropriate, even under variance heterogeneity.}

\section{Algorithmic Variants and Their Computational Complexity} \label{sec:complexity}

\NEW{The computational complexity of transformation trees and forests basically
depends on the variable and split selection performed in every node of
the corresponding trees. In this section, we present several possible
algorithms for the selection of the ``best'' binary split and discuss
corresponding statistical properties and computational complexities. For a
discussion regarding the complexity of random forests we refer to
\cite{Louppe_2015}.}

\NEW{Many prominent tree algorithms, such as CART \citep{Breiman_1984} or C4.5
\citep{Quinlan_1993} evaluate all possible binary splits in all predictor
variables via an exhaustive search. For transformation trees, an exhaustive
search
\begin{eqnarray*}
\hat{\mathcal{B}} = \argmax_{\mathcal{B} \subset \samX} \max_{\parm_1, \parm_2
\in \Theta} \sum_{i = 1}^N I(\rx_i \in \mathcal{B}) \ell_i(\parm_1) + 
\sum_{i = 1}^N I(\rx_i \in \bar{\mathcal{B}}) \ell_i(\parm_2)
\end{eqnarray*}
would require to evaluate the log-likelihood for all possible splits in $O(P
N^2)$. In addition, variable selection based on an exhaustive search would
be biased towards variables
with many potential splits \citep{Kass_1980}. Unbiased recursive
partitioning \citep[for example][]{LohShih1997,Hothorn_Hornik_Zeileis_2006,Zeileis_Hothorn_Hornik_2008} 
separates variable and split selection to address this
bias and to reduce the complexity. Therefore, transformation trees extend
the concept of unbiased recursive partitioning by first selecting the most
important predictor variable by means of a permutation score test and, in a
second step, by finding the best split in this variable as follows (for the sake of simplicity 
we consider the root node only).}

\subsection{Variable Selection} 

\NEW{Transformation trees select the predictor variable with highest association
to the score vector as measured by the $p$-value of a permutation test using
the following procedure:
\begin{enumerate}
\item Compute the maximum likelihood estimator $\hat{\parm}_\text{ML}^N$ in $O(P N)$.
\item Compute the score vector $\s(\hat{\parm}_\text{ML}^N \mid \rY = \ry_i) \in 
      \RR^P$ for each observation $i = 1, \dots, N$ in $O(P N)$.
\item For each predictor variable $j = 1, \dots, J$, compute the linear statistic 
\begin{eqnarray*}
\mT_j = \sum_{i = 1}^N g(x_{ij}) \s(\hat{\parm}_\text{ML}^N \mid \rY = \ry_i)^\top,
\end{eqnarray*}
where $x_{ij}$ is the value of the $j$-th predictor variable for the $i$-th
observation.
The time complexity depends on $g$ and the measurement scale of $X_j$. For a
simple test with high power directed towards linear alternatives $g(x) = x$ is used, with time complexity
$O(P N)$. For a maximally-selected statistic with $g(x) = I(x < x_{i^\prime j}), i^\prime = 1, \dots, N$,
directing high power towards abrupt-change alternatives,
the complexity increases to
$O(P N \log N)$ when the number of potential splits $x_{i^\prime j}$ is allowed to grow
with $N$. 
\item Compute all $P$ corresponding test statistics
\begin{eqnarray*}
\max \left| \frac{\mT_j - \Ex(\mT_j)}{\text{diag}(\V(\mT_j))} \right| \quad \text{or} \quad (\mT_j -
\Ex(\mT_j)) \V(\mT_j)^{-1}(\mT_j - \Ex(\mT_j))
\end{eqnarray*}
in $O(P)$ (best case with $g(x) = x$) or $O(P N)$ (worst case for a maximally selected
statistic) and derive the corresponding $p$-value in $O(1)$. $\Ex(\mT_j)$ and $\V(\mT_j)$
are the conditional expectation and covariance given all admissible
permutations, see \cite{strasserweber1999, Hothorn_Hornik_vandeWiel_2006}. 
Select the variable with lowest $p$-value.
\end{enumerate}}

\NEW{With $g(x) = x$ transformation trees perform the variable selection in $O((J
+ 2) P N)$ instead of the usual $O(J N \log N)$ when an exhaustive search
strategy is employed (for example, in CART).  The test statistic has high
power (and thus the corresponding predictor variable has a high probability
of being selected) when the association to at least one score is linear. In
contrast, the test has low power for $U$-shaped associations, for example.
In such cases, maximally selected statistics with complexity $O(2 P N + J N
\log N)$ have a higher power for detecting such patterns.

Thus, adopting such an inference-based variable selection as opposed to exhaustive
search may also reduce computational complexity. However, unbiasedness is the more
important reason for incorporating the inference-based variable selection
from \cite{Hothorn_Hornik_Zeileis_2006} and \cite{Zeileis_Hothorn_Hornik_2008}
into transformation trees.}

\subsection{Split Selection}

\NEW{Once a predictor variable was selected for splitting, two possible ways for
determining the best split exist. Model-based recursive partitioning 
\citep{Zeileis_Hothorn_Hornik_2008} maximises the log-likelihood over all
possible splits in $O(N^2)$. Transformation trees follow the approach
implemented in conditional inference trees \citep{Hothorn_Hornik_Zeileis_2006}
and select the split based on the score contributions by a maximally
selected statistics of the form
\begin{eqnarray}  \label{fm:maxscore}
\mT_{i^\prime} = \sum_{i = 1}^N I(x_{ij} < x_{i^\prime j}) \s(\hat{\parm}_\text{ML}^N \mid
\rY = \ry_i)^\top, i^\prime = 1, \dots, N
\end{eqnarray}
in $O(P N \log N)$. The best split maximises one of the test statistics
\begin{eqnarray*}
\max \left| \frac{\mT_{i^\prime} -
\Ex(\mT_{i^\prime})}{\text{diag}(\V(\mT_{i^\prime}))} \right| \quad \text{or} \quad
(\mT_{i^\prime} - \Ex(\mT_{i^\prime}))
\V(\mT_{i^\prime})^{-1}(\mT_{i^\prime} - \Ex(\mT_{i^\prime})),
\end{eqnarray*}
for the experiments in Section~\ref{sec:empeval} we used the latter
quadratic form.}

\subsection{Empirical Timings}

\begin{figure}
\begin{center}
\includegraphics{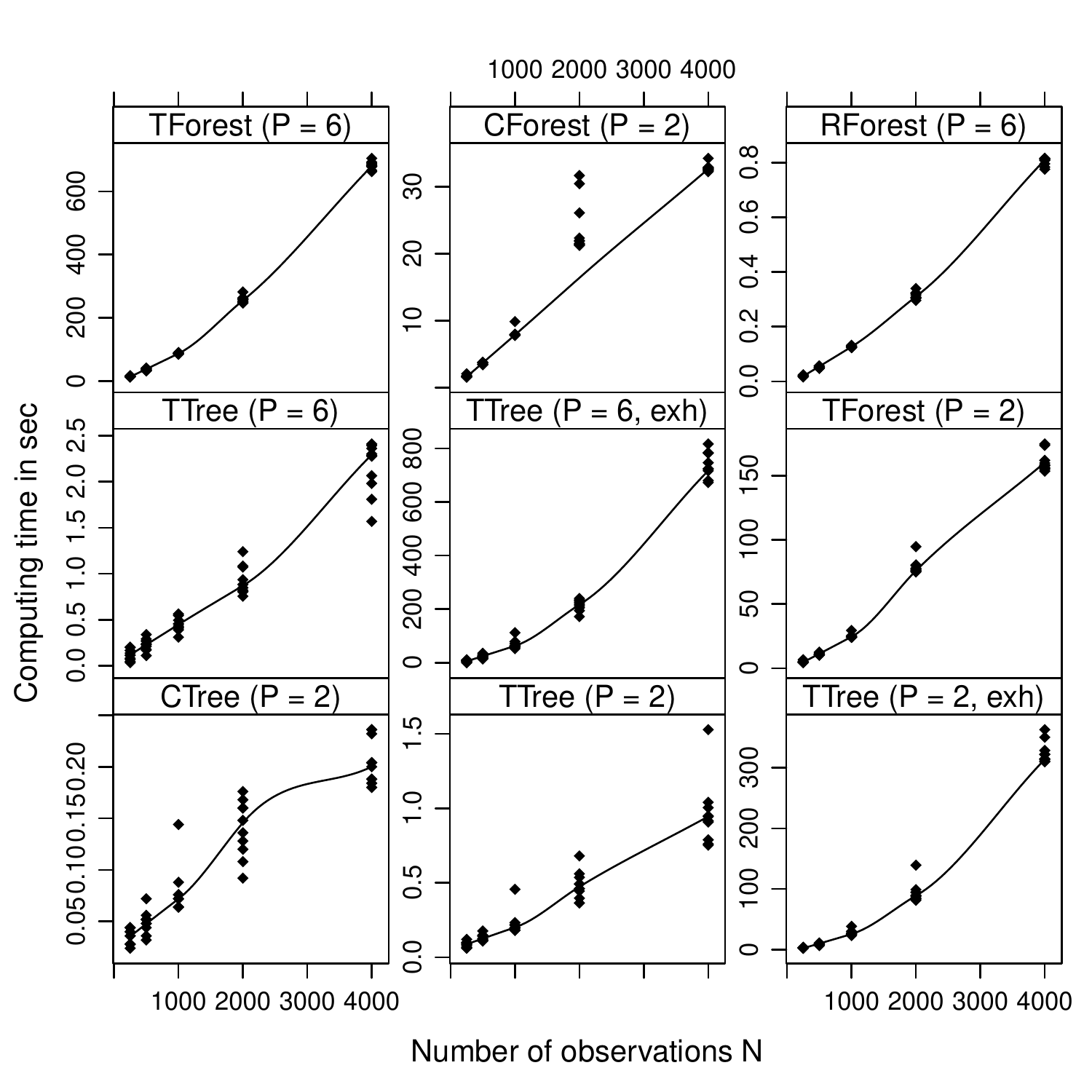}
\caption{Empirical Timings. Run times (in seconds) of all methods compared
in Section~\ref{sec:empeval} for simulation model (\ref{sim:fm}).
Transformation trees with linear ($P = 2$) and non-linear ($P = 6$) transformation
function and with score-based split selection (\ref{fm:maxscore}) and
direct maximisation of the log-likelihood (exh) and corresponding forest
variants are given. \label{timings}}
\end{center}
\end{figure}

\NEW{We compared the run times of the algorithms evaluated in
Section~\ref{sec:empeval} based on model (\ref{sim:fm}) in the informative
low-dimensional setting with varying mean and variance for increasing sample
sizes.  In addition, we added versions of transformation trees with
exhaustive evaluation of all possible splits in the selected variable by
optimising the log-likelihood directly (``exh'').  Figure~\ref{timings}
presents the timings in seconds.}

\begin{figure}
\begin{center}
\includegraphics{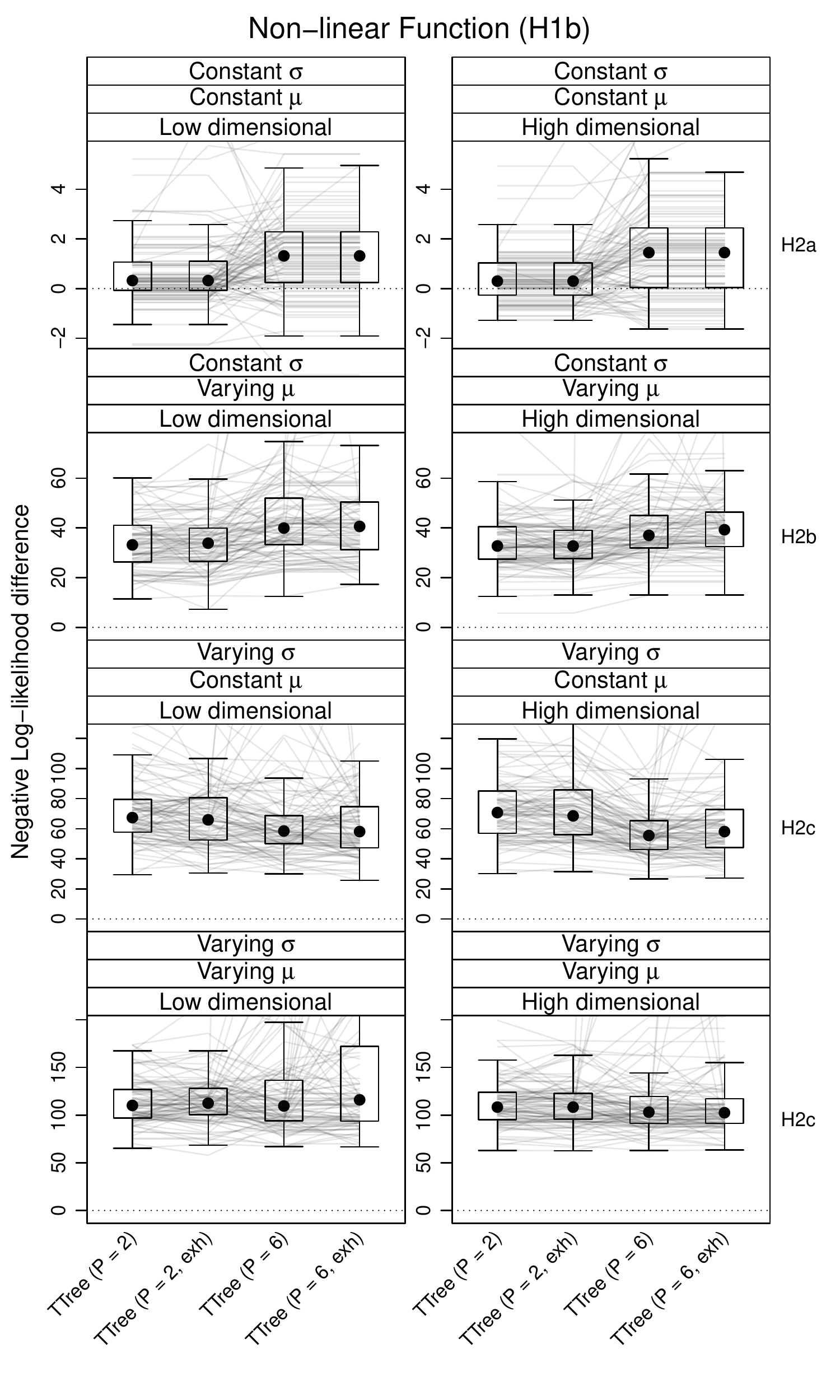}
\caption{Simulation Model (\ref{sim:fm}). Log-likelihood ratios 
for trees and forests in a conditional normal model with
non-linear functions defining mean and variance. 
Transformation trees with linear ($P = 2$) and non-linear ($P = 6$) transformation
function and with score-based split selection (\ref{fm:maxscore}) and
direct maximisation of the log-likelihood (exh) are given. Outliers were not
plotted.
\label{fm-ll-exact}}
\end{center}
\end{figure}

\NEW{CTree and TTree are both based on a linear test statistic with
$g(x) = x$ and a split selection via a maximally selected score statistic
(\ref{fm:maxscore}).  Because steps (1--4) of the variable selection require
most of the time, the total run time is roughly linear in the number of
observations $N$. In contrast, when the split is determined by the maximisation of
the log-likelihood (the ``exh'' option), the split selection dominates
and the increased complexity is visible in the plot.

Note that while the absolute run times differ between algorithms (evident
in the varying y-axis limits), these must not be interpreted as properties
of the algorithms. They just reflect different software design decisions:
For example, the \pkg{randomForest} package \citep{pkg:randomForest} relies
on Breiman and Cutler's original \proglang{Fortran} implementation and is
relatively fast but hard to extend or modify. In contrast, the \pkg{partykit} package
\citep{Hothorn_Zeileis_2015,pkg:partykit} implements a toolbox for recursive
partytioning in high-level \proglang{R} code which is slower
but very flexible and easy to extend. Therefore, transformation trees and forests required
relatively little additional \proglang{R} code ($\sim 500$ lines)
because the infrastructure from \pkg{partykit} and the \pkg{mlt} package for
estimating transformation models \citep{pkg:mlt,vign:mlt.docreg} were straightforward
to reuse.

To check whether trees with and without exhaustive search differ systematically
in their predictive performance, Figure~\ref{fm-ll-exact}
presents a comparison of out-of-sample negative log-likelihoods based on model (\ref{sim:fm}).
Overall, the performance was roughly the same in this situation indicating that the faster score-based approach
was able to identify splits appropriately in this situation.
The empirical complexity of all forest variants was roughly the same, mainly
because the conceptual forest algorithm employed was the same and the only
difference was due to the variable and split selection.}

\subsection{Potential for Optimisation}

\NEW{One potential source of further optimisation is the
ability of transformation models to deal with interval-censored targets.  If
one bins the targets $\ry_i$ into $L + 1$ bins at breaks $-\infty <
\ry_{(1)} < \dots < \ry_{(L)} < \infty$, a model $\Prob_{\rY, \parm}$ of
higher complexity can be fitted by maximising the weighted log-likelihood
for interval-censored observations $\log(\pZ(\h(\ry_{(l)})) - \pZ(\h(\ry_{(l
- 1)})))$ when $\ry_{(l - 1)} < \ry_i \le \ry_{(l)}$.  Evaluation of the
likelihood involves now only $K$ instead of $N$ summands when the data were
tabulated first.  In combination with binned predictor variables,
improvements with respect to computing time and memory
consumption are possible because the linear statistic $\mT_j$ can be
computed based on the contingency table of the binned target and the
binned predictor variable.}

%
%

\section{Discussion}

Transformation forests, as well as the underlying transformation trees, can
be understood as adaptive local likelihood estimators in the rather general
parametric transformation family of distributions.  Owing to possible
interactions and non-linear effects in a ``black-box'' conditional parameter
function $\parm(\rx)$, the resulting conditional distributions of the target
may depend on the predictors in a very general way.  The ability to model
the impact of some predictors on the whole conditional distribution simultaneously,
including its mean and higher moments, is a unique feature of
this novel member of the random forest family.  The likelihood approach
taken here also directly allows the procedures to be applied to randomly
censored or truncated observations.


The algorithmic internals of transformation trees are rooted in conditional
inference trees \citep{Hothorn_Hornik_Zeileis_2006} and model-based
recursive partitioning \citep{Zeileis_Hothorn_Hornik_2008} and inherit the
unbiased variable selection property from these ancestors.  Transformation
forests also allow for unbiased variable importances
\citep{Strobl_Boulesteix_Zeileis_2007}, including the internal handling of
missing predictor variables \citep{Hapfelmeier_Hothorn_Ulm_Strobl_2014}.  An
open-source implementation of transformation trees and transformation forests
based on the \pkg{partykit} add-on package
\citep{Hothorn_Zeileis_2015} to the \textsf{R} system for statistical
computing is available as add-on package \pkg{trtf} \citep{pkg:trtf}, see
Appendix~``Computational Details''.

Within the theory of adaptive local likelihood estimation, alternative
choices for parametric models (via their likelihood contributions
$\ell_i(\parm)$) and weights $w_i^N(\rx)$ are possible.  In the context of
personalised medicine or personalised marketing, one is interested in the
dependency of some treatment effect $\beta$ on predictors $\rx$.  Random
forest-type algorithms are a promising tool for modelling complex effects of
predictors on such a treatment parameter \citep{Foster_Taylor_Ruberg_2011,
Seibold_Zeileis_Hothorn_2015, Wager_Athey_2015,
Seibold_Zeileis_Hothorn_2016}.  In the framework presented here,
implementation of such a strategy only requires the specification of a
distribution $\Prob(\rY \le \ry \mid \text{treated}) = \pZ(\basisy(\ry)^\top
\parm + \beta)$ for treated and $\Prob(\rY \le \ry \mid \text{untreated}) =
\pZ(\basisy(\ry)^\top \parm)$ for untreated observations.  The model, and
therefore also the treatment effect $\beta$, can then be partitioned or
aggregated by transformation trees and transformation forests leading to a
random forest estimate $\hat{\beta}^N_\text{Forest}(\rx)$ of the conditional
treatment effect $\beta(\rx)$ in addition to
$\hat{\parm}^N_\text{Forest}(\rx)$.  \NEW{Breiman and Cutler's random
forests were empirically shown to be insensitive to
changes in treatment effects $\beta$ \citep[in comparison to adaptive local
likelihood estimation of very simple parametric models, such as logistic or
Weibull regression,][]{Seibold_Zeileis_Hothorn_2016}.  This corresponds to
the empirical findings reported in Section~\ref{sec:empeval} showing an
insensitivity of Breiman and Cutler's random forests to changes in the variance of a conditional
normal distribution.  Transformation trees and forests, in contrast, were
specifically designed to detect such distributional changes by an assessment
of parameter stability.  This property extends to additional parameters in
more complex transformation models featuring predictor-varying effects
$\parm(\rx)$ and $\shiftparm(\rx)$ of the form \begin{eqnarray*} \Prob(\rY
\le \ry \mid \rX = \rx, \rU = \ru) = \pZ(\basisy(\ry)^\top \parm(\rx) +
\ru^\top \shiftparm(\rx)) \end{eqnarray*} which describe the most general
model class associated with transformation trees and forests.  Unlike the
local normal linear models studied in \cite{Bloniarz_Wu_Yu_2016}, the
general framework proposed here allows for varying linear effects
$\shiftparm(\rx)$ of additional predictor variables $\ru$ (as, for example,
treatment effects in randomised trials or a priori known confounders in
observational studies).  Beyond this additional modelling flexibility and
unlike Breiman and Cutler's random forests, transformation trees and forests also allow for all
types of target variables under all forms of random censoring or truncation
\citep[an overview on known and unknown models from this class is available
from Section~4.3 and Table~1 in][]{Hothorn_Moest_Buehlmann_2016}.  When
parameter estimates for such a transformation model are already given (\ie
when some elements of $\shiftparm(\rx)$ are available from an external
source and shall be kept fix, for example some established treatment
effect), one could use transformation forests to estimate a deviation from
this initial model.  An already existing transformation function can be used
as an offset in the likelihood $\ell_i$, such that the forest conditional
parameter function excludes these existing effects.}

A more general understanding of weights could be derived from the notion of applying a
distance measure $d$ to two distributions 
$\hat{\Prob}_{\rY \mid \rX = \rx} = \hat{\Prob}_{\rY, \hat{\parm}_t(\rx)}$ and 
$\hat{\Prob}_{\rY \mid \rX = \rx_i} = \hat{\Prob}_{\rY, \hat{\parm}_t(\rx_i)}$
obtained from the $t$-th tree. Based on this distance, an alternative weight could
be defined by
\begin{eqnarray*}
w_{\text{Forest}, i}^N(\rx) = \sum_{t = 1}^T \left(1 -
d\left(\hat{\parm}_{\text{Tree}, t}^N(\rx), \hat{\parm}_{\text{Tree}, t}^N(\rx_i)\right)
\right)
\end{eqnarray*}
for example using the Kullback-Leibler divergence for continuous distributions
\begin{eqnarray*}
d_\text{KL}(\parm_1, \parm_2) = \int \dY(\ry \mid \parm_1) \log\left(\frac{\dY(\ry \mid\parm_1)}{\dY(\ry \mid \parm_2)}\right) \,d\ry
\end{eqnarray*}
(after standardisation to the unit interval). This weight takes the conditional
distribution in two terminal nodes of a tree into account, rather than just treating
them as ``somehow different'' in the way of nearest neighbour weights.

\NEW{The empirical evaluation of transformation trees and transformation
forests for censored targets 
\citep[and comparison to a new competitor which is based on splits
maximising the integrated absolute difference between conditional survivor
curves, recently published by][]{Moradian_Larocque_Bellavance_2016} as well as
the evaluation of the quality of likelihood-based permutation variable
importance (including the conditional variable importance) for variable selection,
of the model-based bootstrap for variability assessment, and of the likelihood-ratio test
are ongoing research projects.}




\bibliography{mlt,packages}


\begin{appendix}

\section*{Appendix}

\subsection*{Computational Details}

A reference implementation of \underline{tr}ansformation \underline{t}rees
and transformation \underline{f}orests is available in the \pkg{trtf}
package \citep{pkg:trtf}.  This package was built on top of the infrastructure
packages \pkg{partykit} \citep{Hothorn_Zeileis_2015,pkg:partykit} and 
\pkg{mlt} \citep{pkg:mlt,vign:mlt.docreg}. Conditional inference
trees and forests were fitted using package \pkg{partykit}. Quantile regression
forests were computed by the \pkg{quantregForest} package
\citep{pkg:quantregForest}.  The reference implementation 
of Breiman's and Cutlers random forests in the \pkg{randomForest} package \citep{pkg:randomForest} was used.
All computations were performed using
\textsf{R} version 3.4.3
\citep{R}.

For the empirical evaluation in Section~\ref{sec:empeval}, all
non-linear transformation models were based on transformation functions parameterised
in terms of Bernstein polynomials of order five, \ie with six parameters,
and $\pZ = \Phi$. 
Log-likelihoods were optimised under monotonicity constraints using a
combination of augmented Lagrangian minimisation and spectral projected
gradients.  Unbiased trees, including transformation trees, stopped
internally when the minimum Bonferroni-adjusted $p$-value was larger than
$0.05$.  No such internal stopping was applied in conditional inference or
transformation forests.  Subsampling
of $.632 N$ observations was used for all random forest-types.  
The minimum number of observations necessary for splitting 
(\texttt{minsplit} in \pkg{partykit} and \texttt{nodesize} in
\pkg{randomForest}) was $25$ for all forest types in the simulation
experiments. 

Data from the Swiss Health Survey 2012 can be obtained from the Swiss
Federal Statistics Office (Email: \url{sgb12@bfs.admin.ch}).  Data is available
for scientific research projects, and a data protection application form
must be submitted.  More information can be found here
\url{http://www.bfs.admin.ch/bfs/de/home/statistiken/gesundheit/erhebungenSupplementary}.
The code used for producing the results for the body mass illustration can be
evaluated on a smaller artificial data set sampled from the transformation
forest by running \texttt{demo("BMI")} from the \pkg{trtf} package
\citep{pkg:trtf}; Figure~\ref{fig:exIntro} is regenerated by
\texttt{demo("QRF")}. The simulation results presented in this paper can be reproduced using the
files in \texttt{system.file("sim", package = "trtf")}.

\subsection*{Additional Results: Empirical Evaluation}

\subsubsection*{Additional Evaluation of Tree-Structured Conditional
Parameter Function (H1a)}

\begin{figure}
\begin{center}
\includegraphics{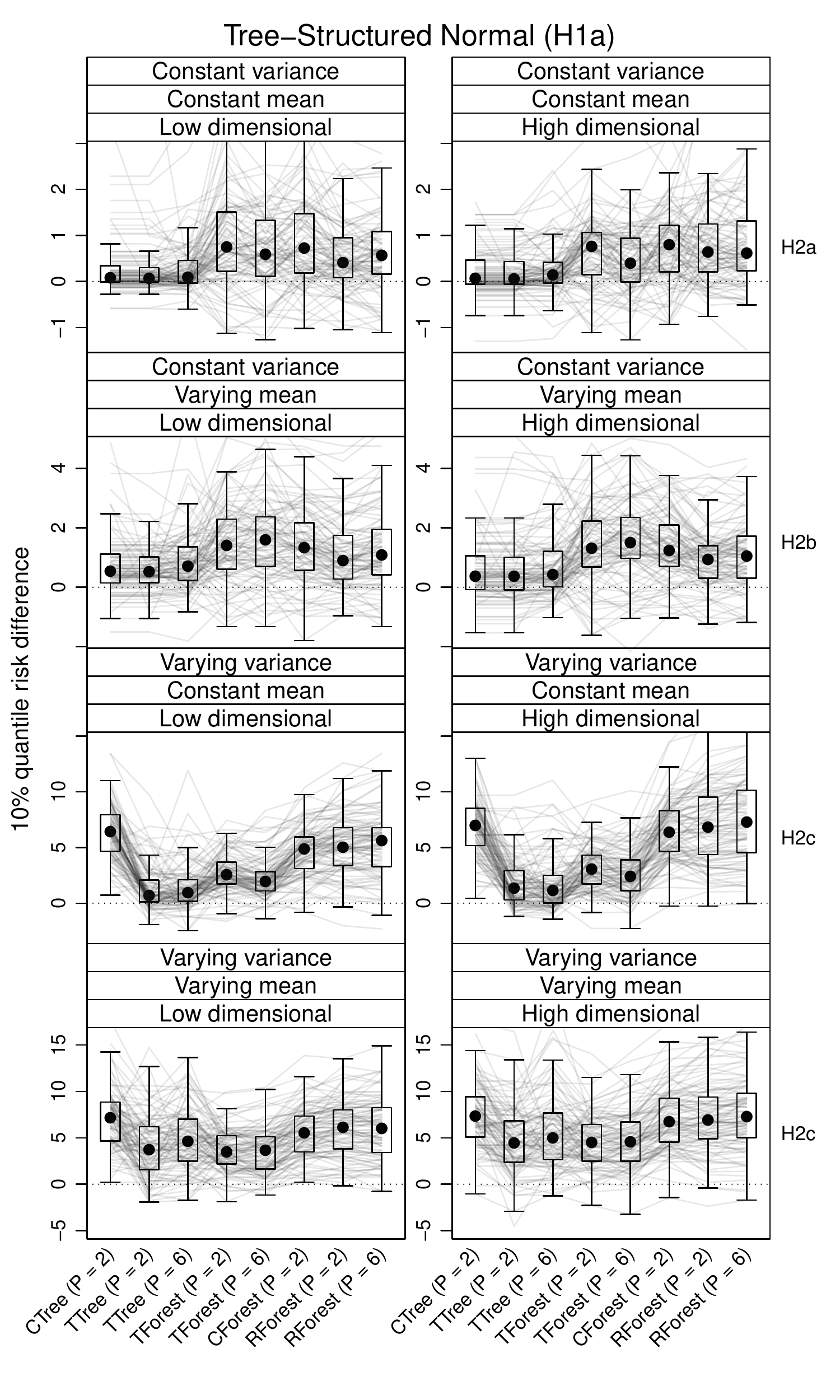}
\caption{Simulation Model (\ref{sim:2d}). $10\%$ quantile risk differences
  for trees and forests in a conditional normal model with
  non-linear functions defining mean and variance. 
  The quantile risk difference was computed
  as the out-of-sample check risk of each competitor
  minus the check risk of the true data generating process.
\label{2d-L1}}
\end{center}
\end{figure}

\begin{figure}
\begin{center}
\includegraphics{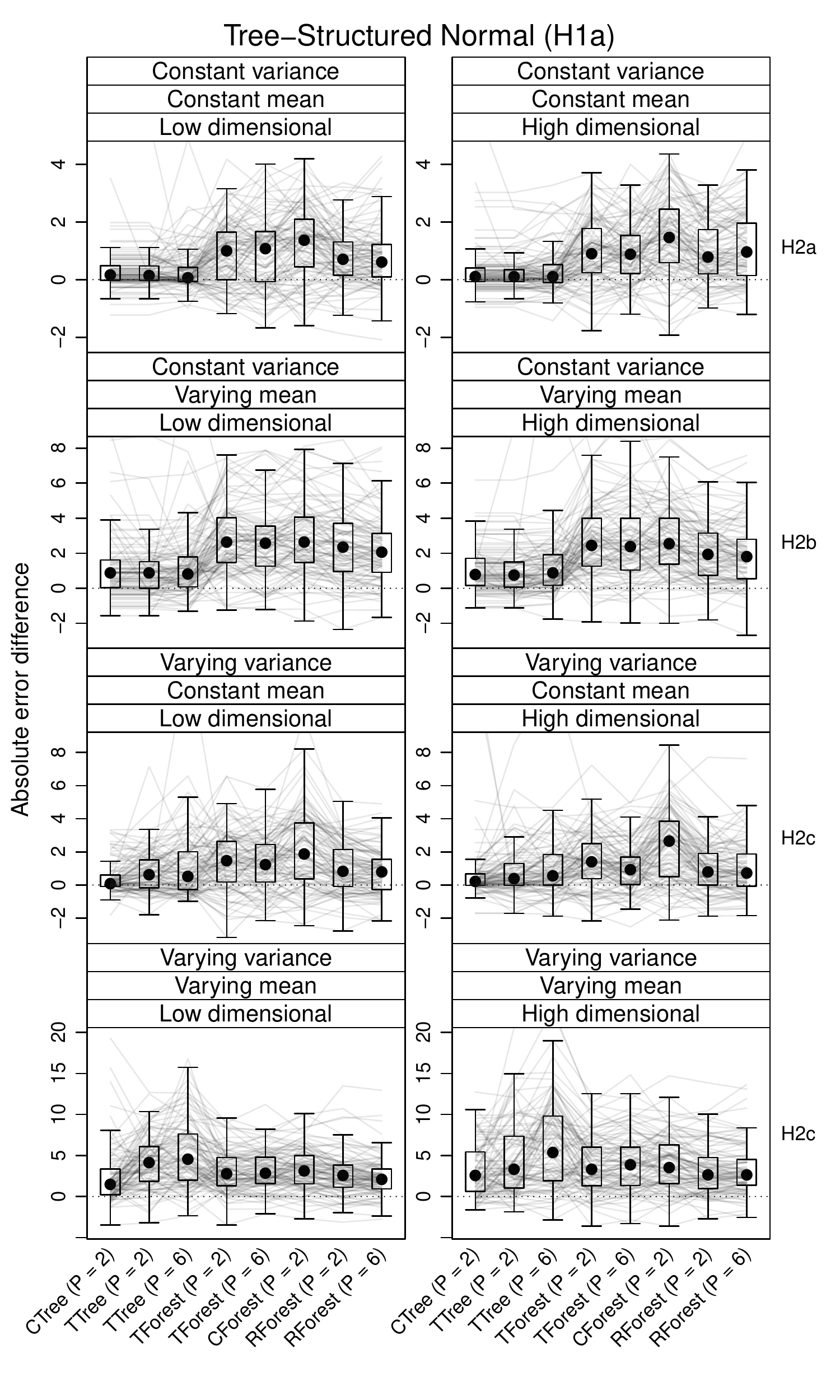}
\caption{Simulation Model (\ref{sim:2d}): Absolute error differences
  for trees and forests in a conditional normal model with
  non-linear functions defining mean and variance. 
  The absolute error difference was computed
  as the absolute error of each competitor
  minus the absolute error of the true data generating process.
\label{2d-L5}}
\end{center}
\end{figure}

\begin{figure}
\begin{center}
\includegraphics{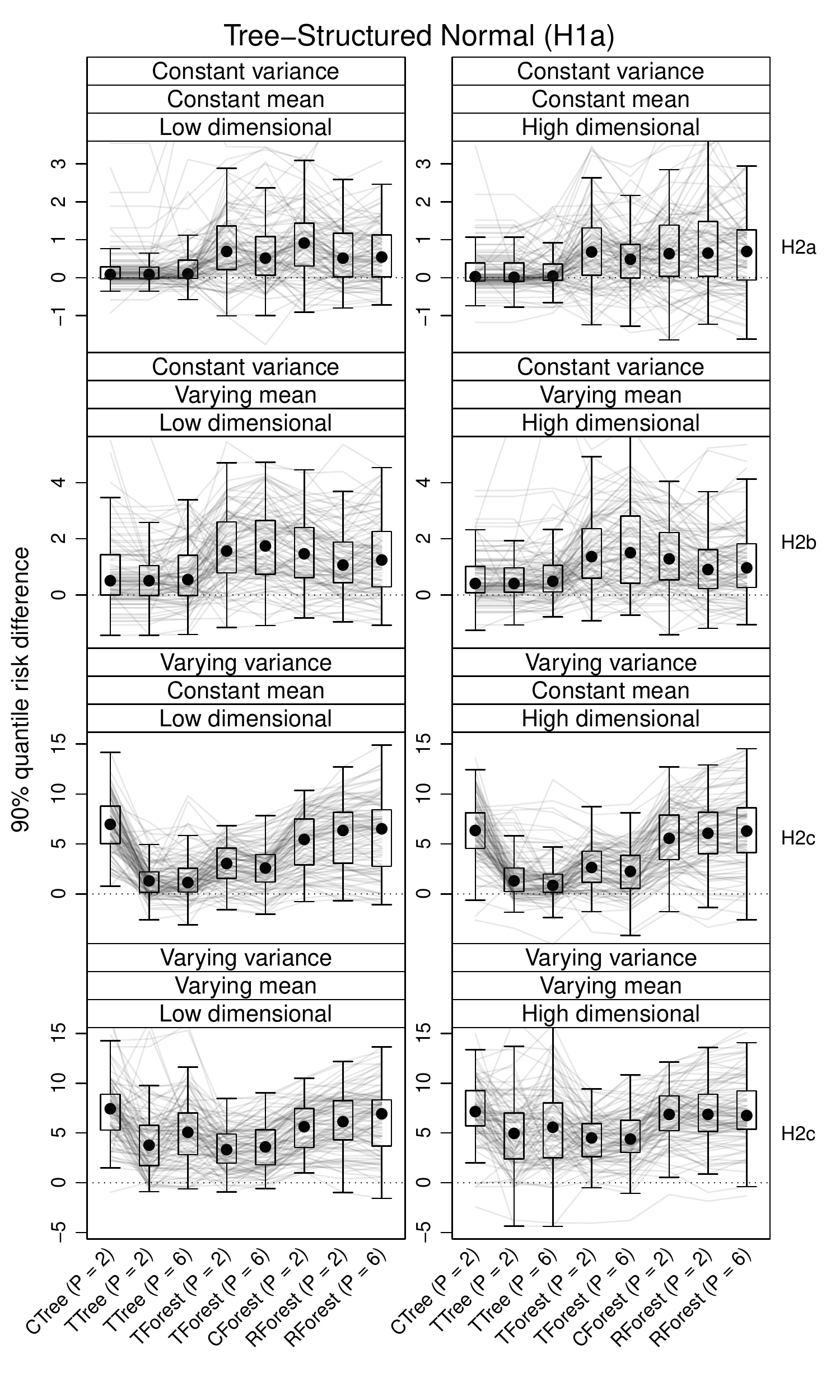}
\caption{Simulation Model (\ref{sim:2d}): $90\%$ quantile risk differences
  for trees and forests in a conditional normal model with
  non-linear functions defining mean and variance. 
  The quantile risk difference was computed
  as the out-of-sample check risk of each competitor
  minus the check risk of the true data generating process.
\label{2d-L9}}
\end{center}
\end{figure}

\begin{figure}
\begin{center}
\includegraphics{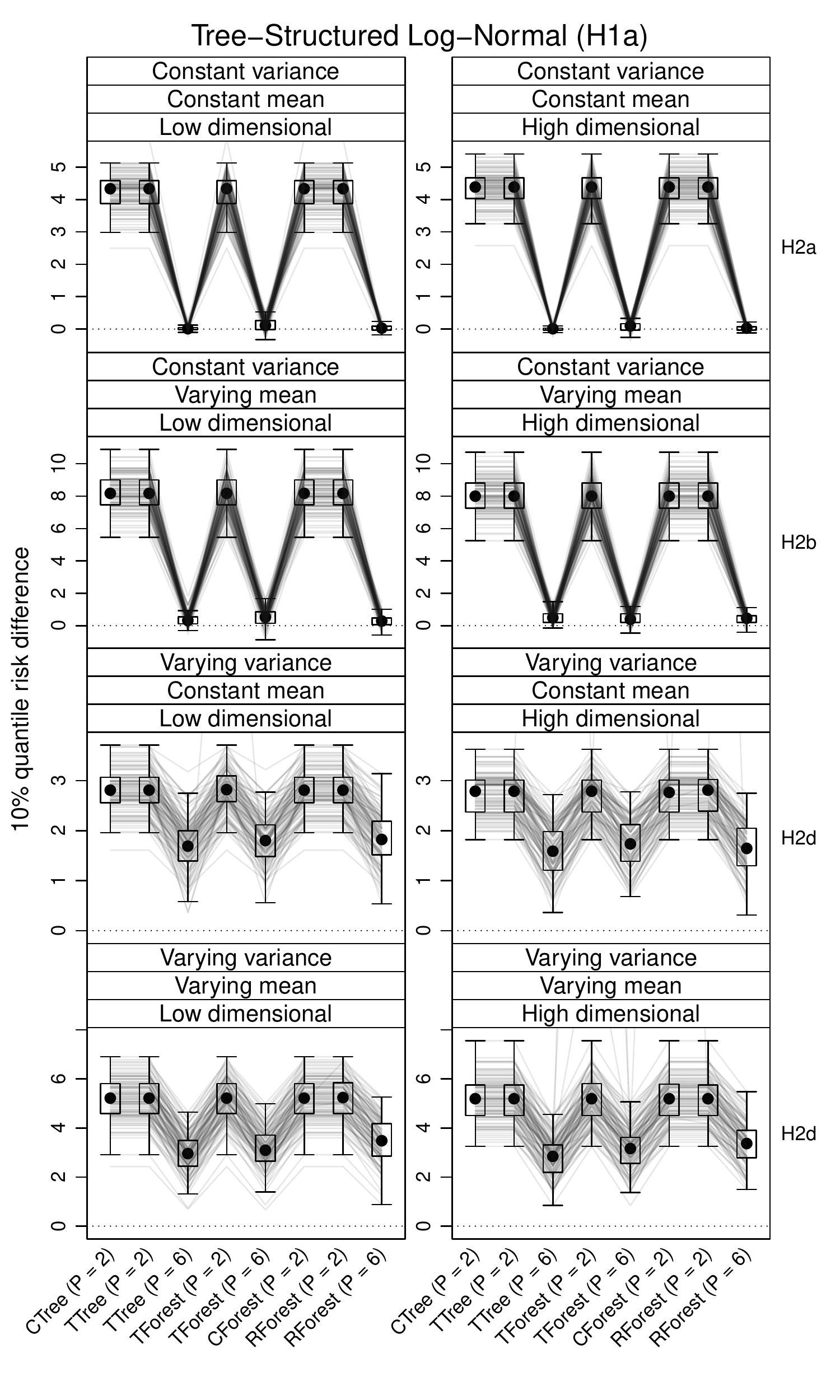}
\caption{Simulation Model (\ref{sim:ln-2d}). $10\%$ quantile risk differences
  for trees and forests in a conditional log-normal model with
  non-linear functions defining mean and variance. 
  The quantile risk difference was computed
  as the out-of-sample check risk of each competitor
  minus the check risk of the true data generating process. \label{ln-2d-L1}}
\end{center}
\end{figure}

\begin{figure}
\begin{center}
\includegraphics{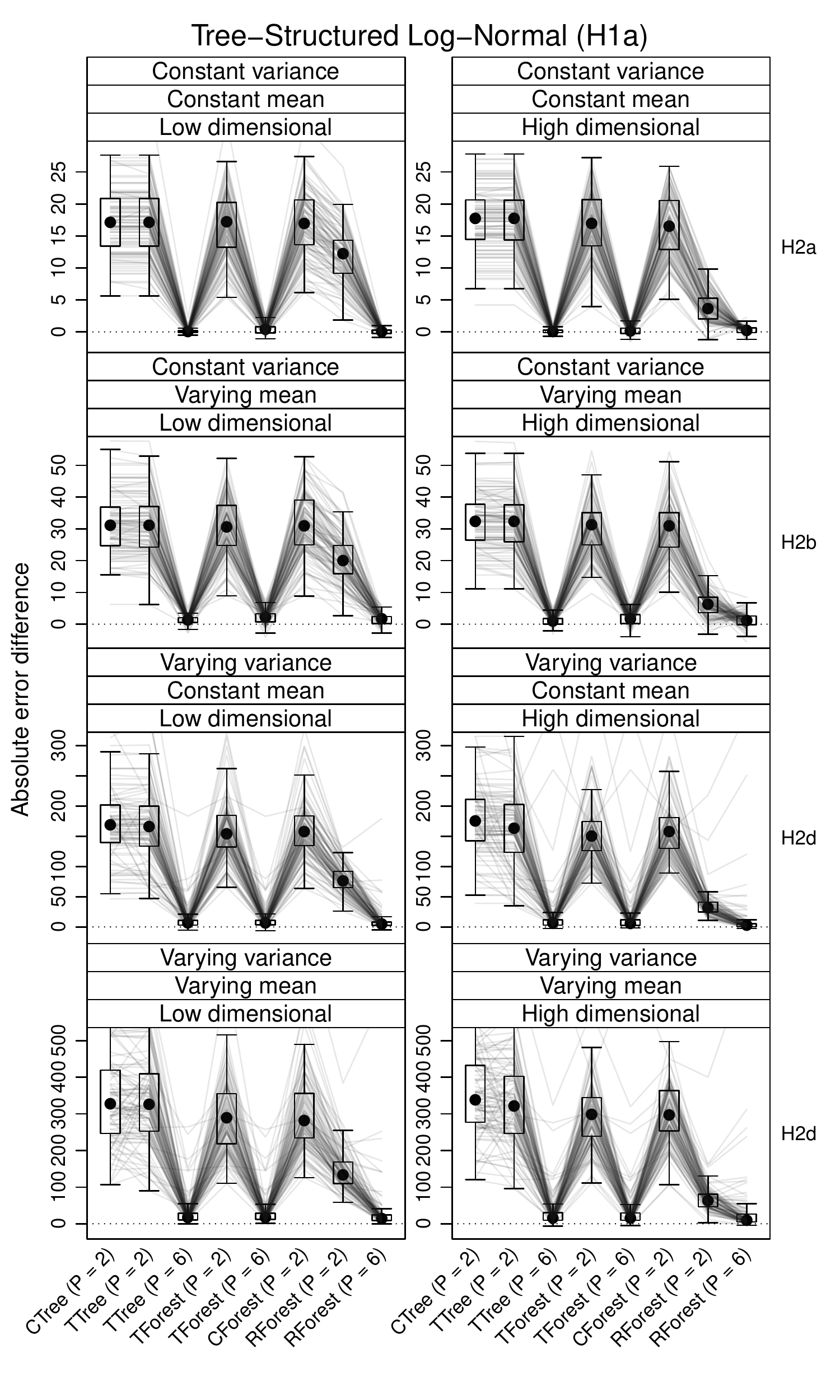}
\caption{Simulation Model (\ref{sim:ln-2d}): Absolute error differences
  for trees and forests in a conditional log-normal model with
  non-linear functions defining mean and variance. 
  The absolute error difference was computed
  as the absolute error of each competitor
  minus the absolute error of the true data generating process.
\label{ln-2d-L5}}
\end{center}
\end{figure}

\begin{figure}
\begin{center}
\includegraphics{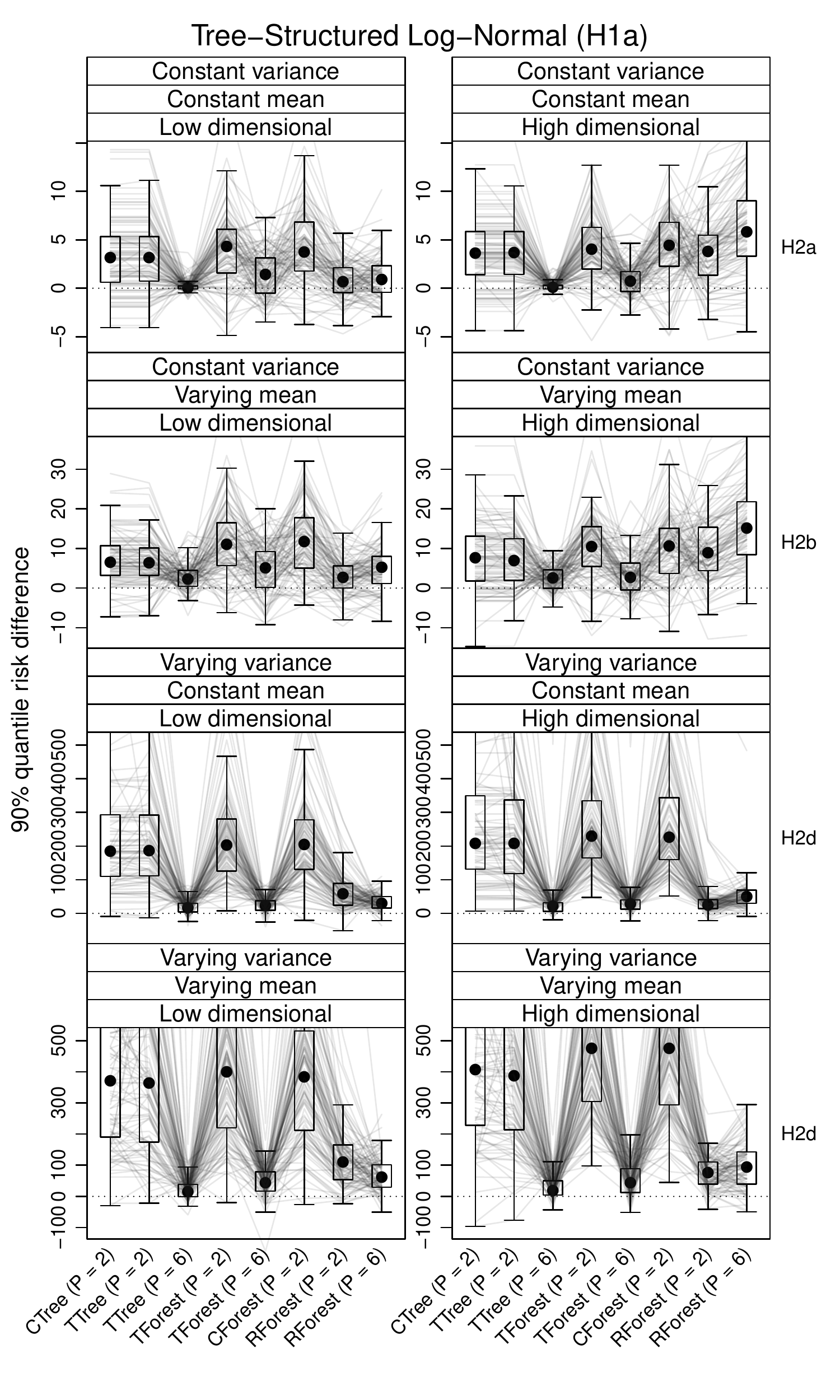}
\caption{Simulation Model (\ref{sim:ln-2d}): $90\%$ quantile risk differences
  for trees and forests in a conditional log-normal model with
  non-linear functions defining mean and variance.  
  The quantile risk difference was computed
  as the out-of-sample check risk of each competitor
  minus the check risk of the true data generating process.
\label{ln-2d-L9}}
\end{center}
\end{figure}

\subsubsection*{Additional Evaluation of Non-Linear Conditional Parameter Function (H1b)}

\begin{figure}
\begin{center}
\includegraphics{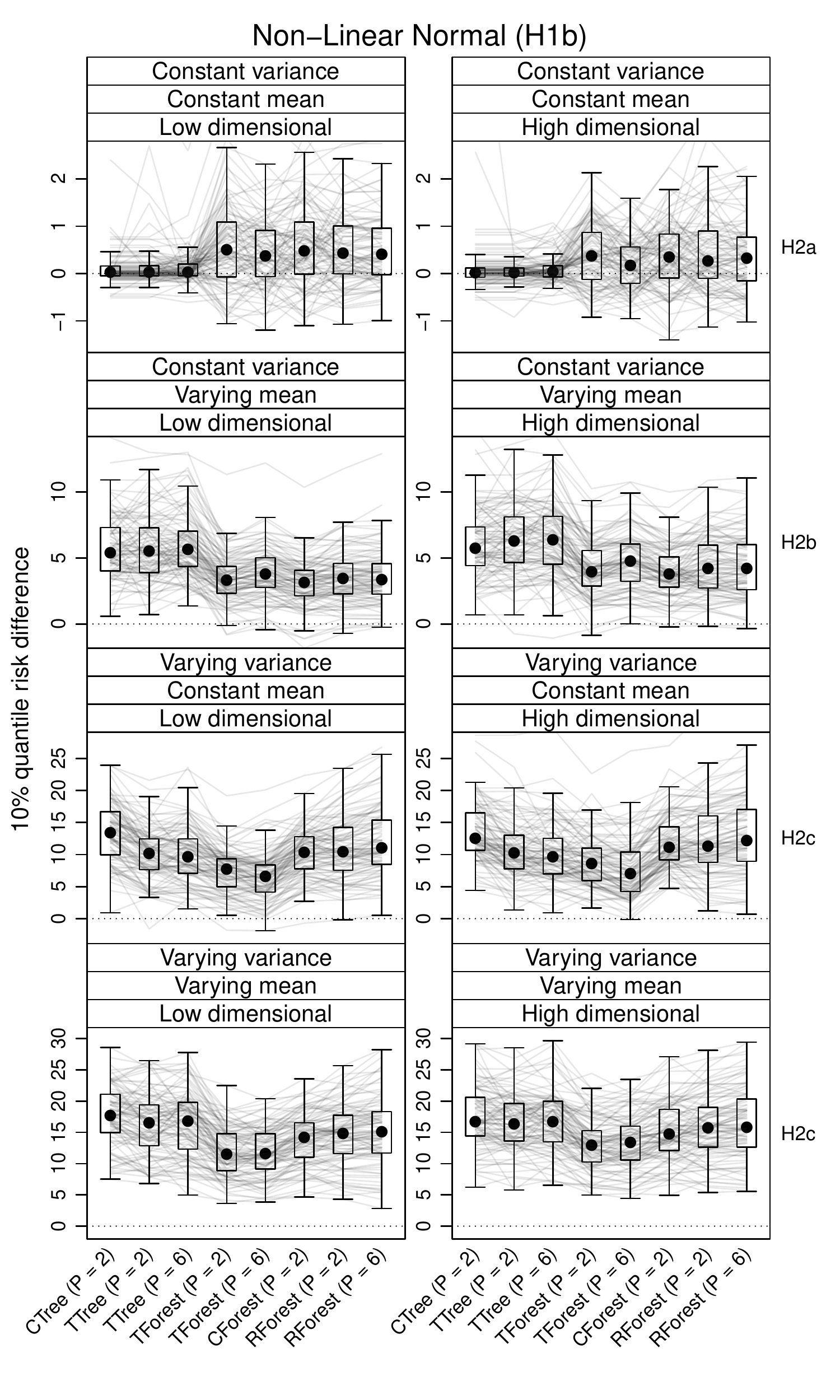}
\caption{Simulation Model (\ref{sim:fm}). $10\%$ quantile risk differences
  for trees and forests in a conditional normal model with
  non-linear functions defining mean and variance. 
  The quantile risk difference was computed
  as the out-of-sample check risk of each competitor
  minus the check risk of the true data generating process.
\label{fm-L1}}
\end{center}
\end{figure}

\begin{figure}
\begin{center}
\includegraphics{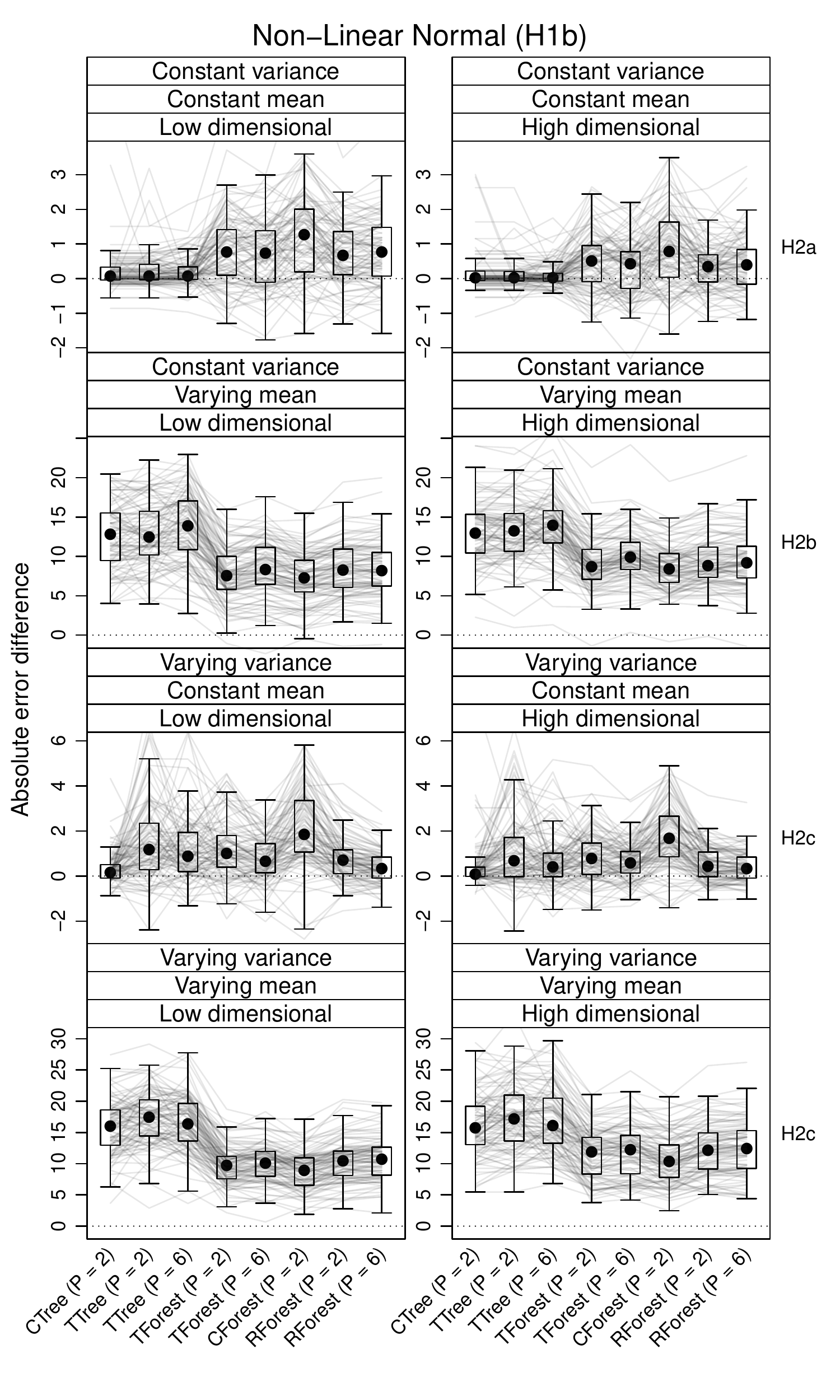}
\caption{Simulation Model (\ref{sim:fm}): Absolute error differences
  for trees and forests in a conditional normal model with
  non-linear functions defining mean and variance. 
  The absolute error difference was computed
  as the absolute error of each competitor
  minus the absolute error of the true data generating process.
\label{fm-L5}}
\end{center}
\end{figure}

\begin{figure}
\begin{center}
\includegraphics{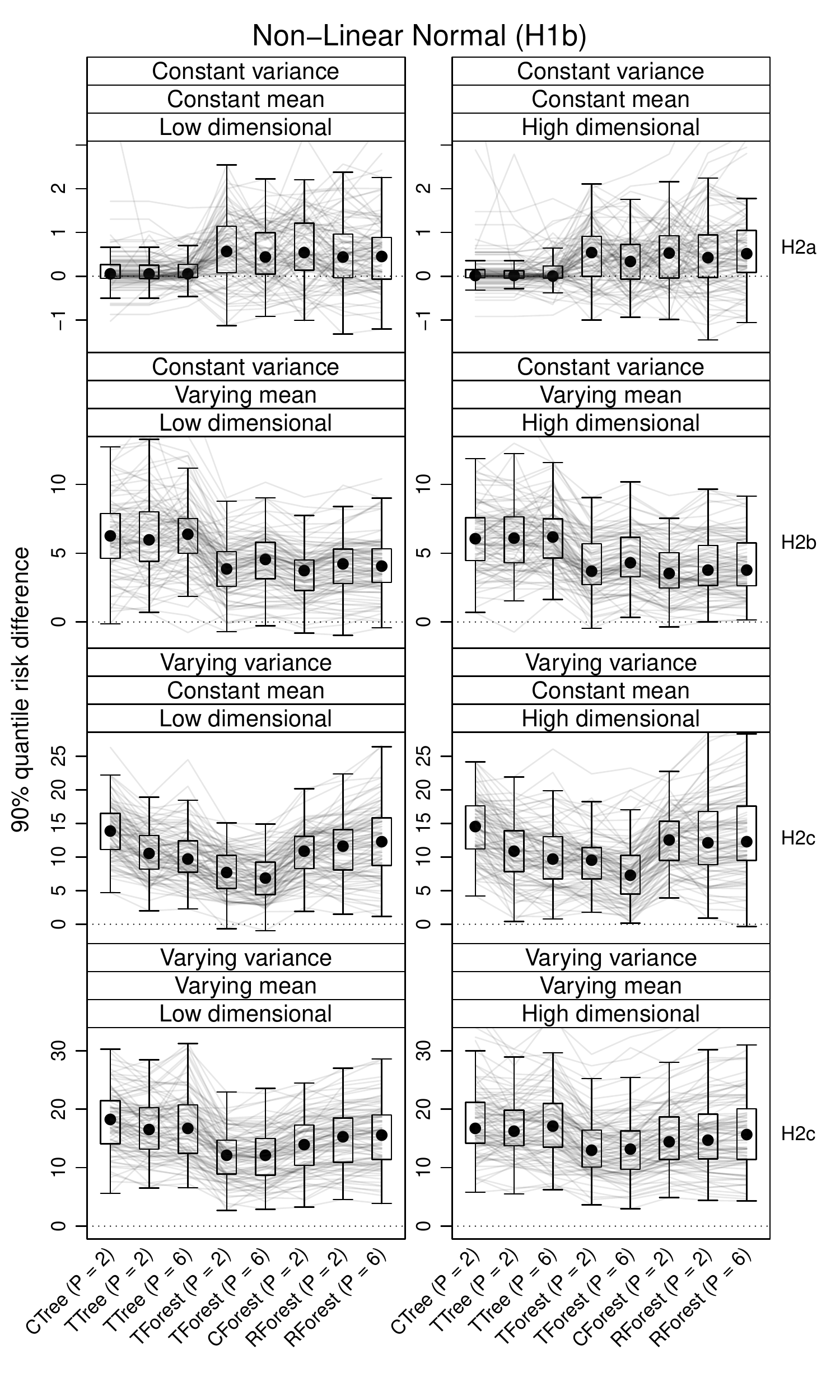}
\caption{Simulation Model (\ref{sim:fm}): $90\%$ quantile risk differences
  for trees and forests in a conditional normal model with
  non-linear functions defining mean and variance. 
  The quantile risk difference was computed
  as the out-of-sample check risk of each competitor
  minus the check risk of the true data generating process.
\label{fm-L9}}
\end{center}
\end{figure}

\begin{figure}
\begin{center}
\includegraphics{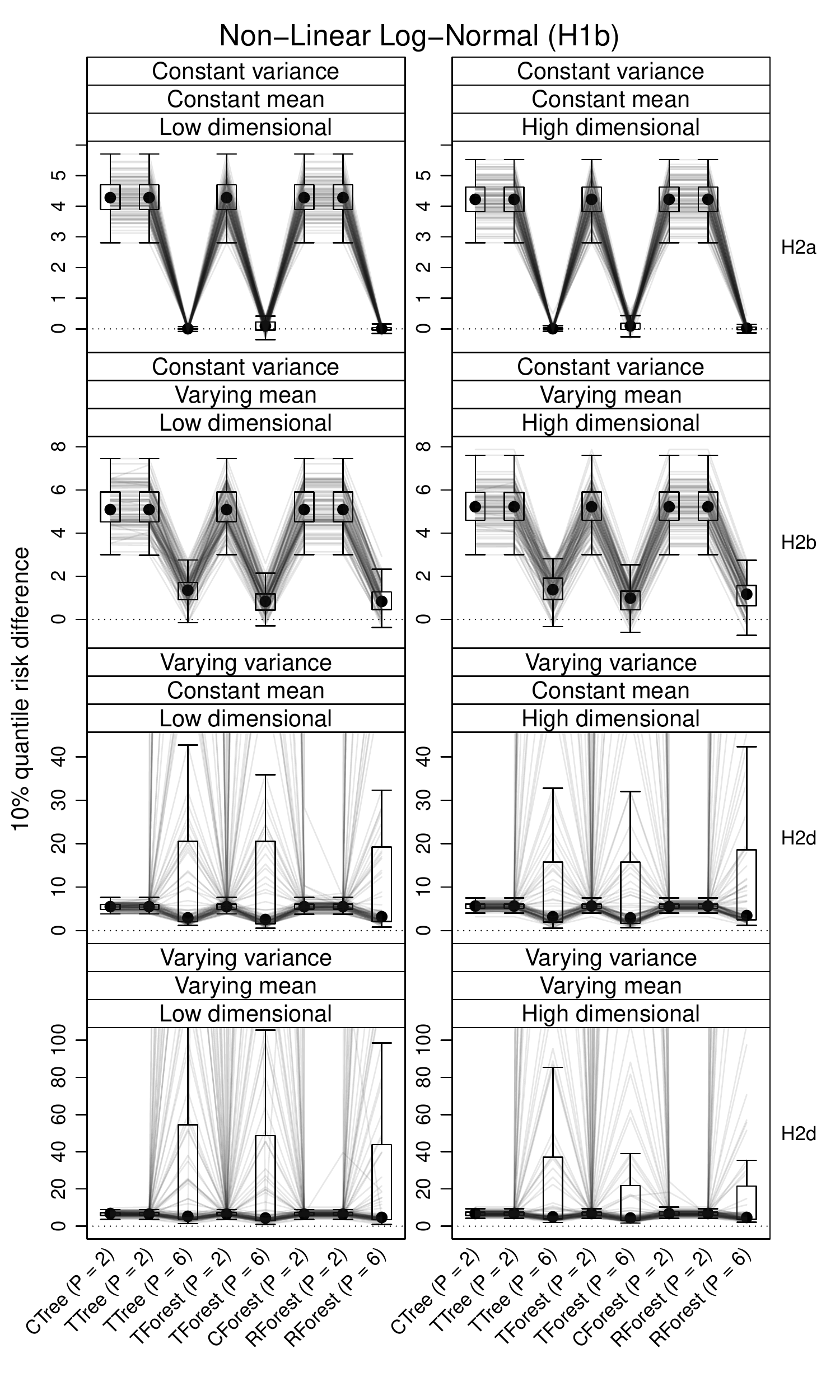}
\caption{Simulation Model (\ref{sim:ln-fm}). $10\%$ quantile risk differences
  for trees and forests in a conditional log-normal model with
  non-linear functions defining mean and variance.  
  The quantile risk difference was computed
  as the out-of-sample check risk of each competitor
  minus the check risk of the true data generating process.
\label{ln-fm-L1}}
\end{center}
\end{figure}

\begin{figure}
\begin{center}
\includegraphics{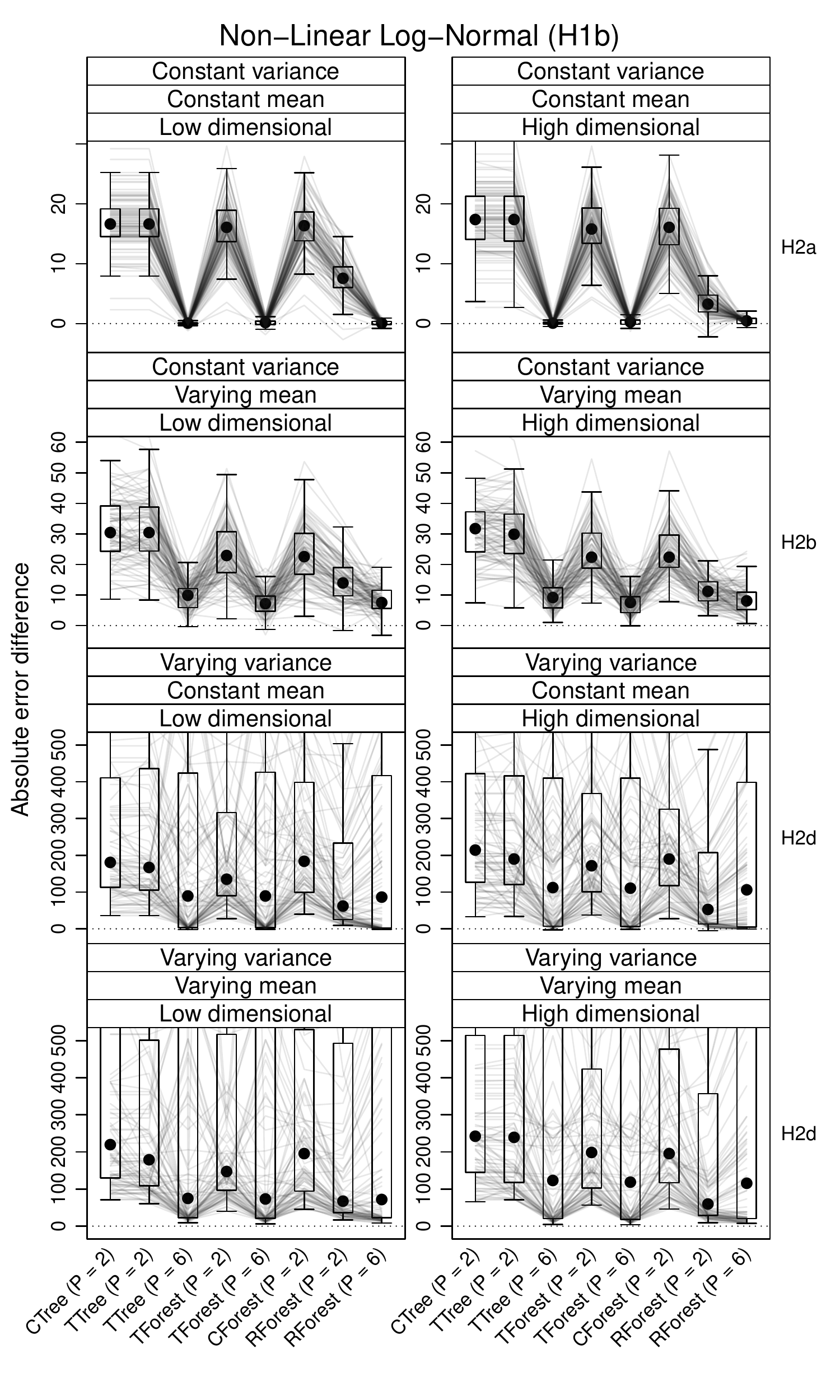}
\caption{Simulation Model (\ref{sim:ln-fm}): Absolute error differences
  for trees and forests in a conditional log-normal model with
  non-linear functions defining mean and variance.  
  The absolute error difference was computed
  as the absolute error of each competitor
  minus the absolute error of the true data generating process.
\label{ln-fm-L5}}
\end{center}
\end{figure}

\begin{figure}
\begin{center}
\includegraphics{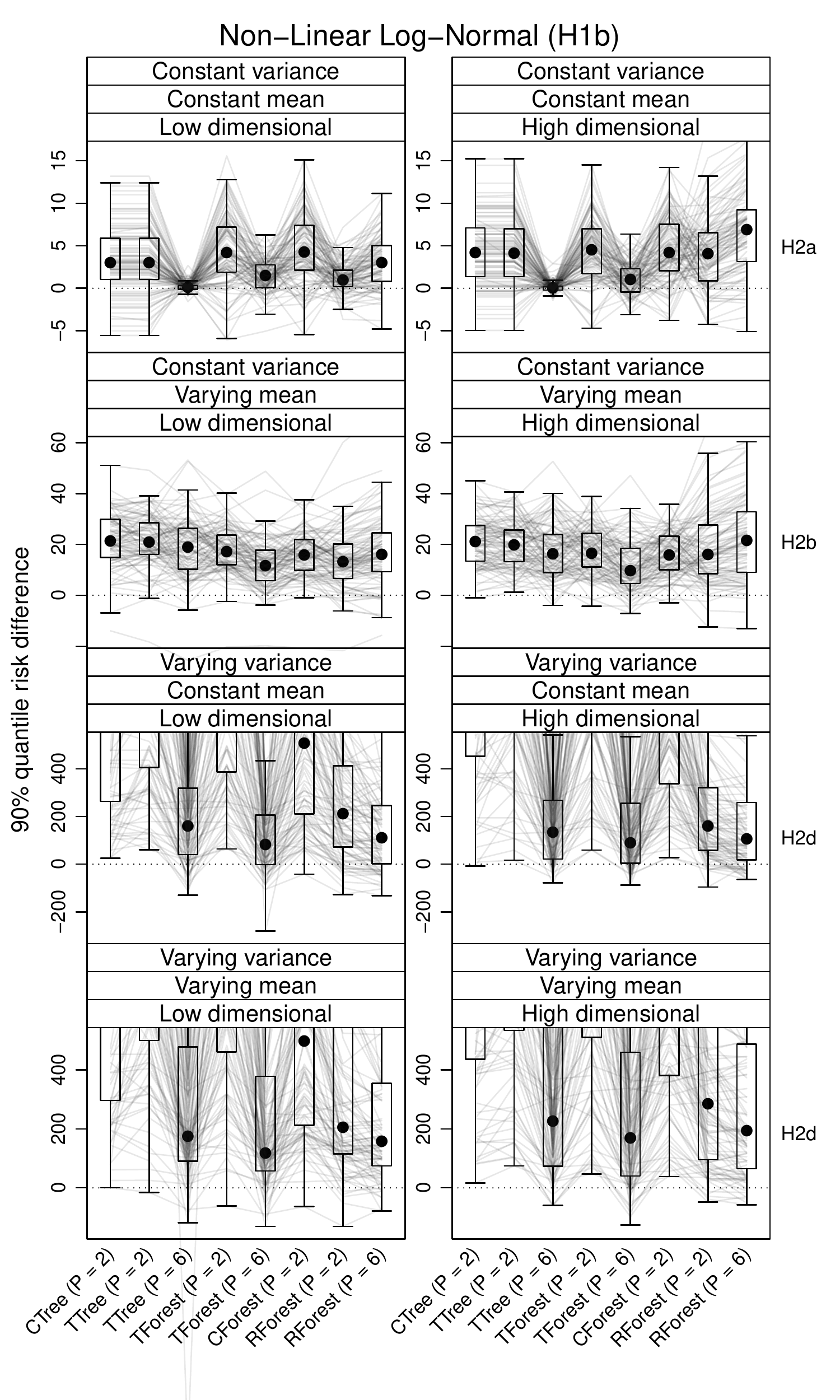}
\caption{Simulation Model (\ref{sim:ln-fm}): $90\%$ quantile risk differences
  for trees and forests in a conditional log-normal model with
  non-linear functions defining mean and variance. 
  The quantile risk difference was computed
  as the out-of-sample check risk of each competitor
  minus the check risk of the true data generating process.\label{ln-fm-L9}}
\end{center}
\end{figure}

\subsection*{Review History: Version 1 by Journal 1 (January 2017--May 2017)}

\noindent Review for version 1 (\url{https://arxiv.org/abs/1701.02110v1}).
Comments by referees are printed in \textit{italics}, replies by the authors
in plain text.

\subsubsection*{Handling Editor}

\noindent
\textit{
Both reports find the paper interesting in that it combines
partitioning methods with local likelihood estimation. However, they also
raise very good questions about the paper. I have also
taken a look at the paper myself. I highlight the important points/issues 
below, combining the reports' comments and my own:}

\begin{enumerate}
\item \textit{Computational complexity of the proposed method, especially in high-dimension, needs to be
              addressed.} \label{AE1}

We added a new Section~8 containing a detailed description of the variable
and split selection methods in transformation trees. This section also discusses the
statistical properties and the computational complexity of these variants.
The theoretical findings are supported by an analysis of the empirical
runtimes.

\item  \textit{Reproducibility of the experimental results is unclear.
               That is, the simulation set-ups
               are not clear enough for a reader to reproduce the results by
               writing their own codes. Please make the results reproducible.}

GPL-2 licensed open-source code implementing transformation trees and
forests is available from CRAN since 2018-01-08 (package \textbf{trtf}). The revision now contains a
link to the source code for the artificial simulation experiments. Reproducibility 
material for the body weight application is published with Hothorn (2018). These issues
are discussed in Section~``Computational Details''.

\item \textit{A much clearer description is needed on how your proposed method is 
              built on existing works in the literature, and please cite all 
              appropriate papers (the second report mentions a few more).}

We agree that the description of unbiased model-based recursive partitioning
was a little too opague in the first version. We comment on statistics used
for variable and split selection in unbiased model-based recursive
partitioning (and thus also by transformation
trees) now in much more detail in our new Section~8.

\item \textit{More explanation and evidence on
              why a new method like yours is needed too, given the many related method --
              I understand that interpretability is a motivation, did you illustrate
              clearly the interpretability of your new method in the data
              example?}

We believe that the new simulation experiments support our main hypothesis that
none of the existing tree or forest algorithms is able to detect
distributional changes unrelated to the mean. Transformation trees and
forests fill this gap. This fact is prominently stated in the introduction
and throughout the manuscript and motivated the design of the new simulation
experiments. The new illustration regarding body weight
gives an (as we hope) easy to understand impression of how transformation
trees and forests detect and represent distributional changes in higher
moments of the conditional distribution of the target given predictor
variables.

\item \textit{Computational complexity comparisons to representatives of the
              existing methods, including in terms of running times of your
              method and others in the experimental results section.}

See point \ref{AE1}.

\item \textit{Since RF is commonly used in practice rather than the
              model-based partitioning method, it is necessary to use one
              version of RF, for example, the one by Breiman, in place of
              the model-based partitioning method, to generate weights used in
              the proposed method.  Comparisons to this RF-based version of
              the proposed methods in terms of computational complexity
              (running time) and prediction and confidence interval metrics
              are needed.}

We agree that most readers will be interested in a direct comparison with
Breiman and Cutler's randomForests. The new simulation experiments directly compare
transformation trees and forests as well as conditional inference trees and
forests with this reference implementation of random forests. We followed the
suggestion to estimate a conditional transformation model based on 
weights extracted from Breiman and Cutler's randomForests. Therefore, a direct
comparison to transformation forests on the same scale is now presented.
Prediction error is assessed based on the out-of-sample log-likelihood and
prediction intervals are evaluated based on the $10\%$ and $90\%$ check
risk. Theoretical and empirical runtimes are studied in our new Section~8.

\item \label{AE:boot}
\textit{Moreover, I am not convinced that model-based bootstrap is a
              good method to use for uncertainty measures, especially in
              high-dim and when the data generating model is misspecified or
              is a mis-match for the parametric models used in your method.}

We did not evaluate the quality of the model-based bootstrap (Section~5.3)
nor size and power of the likelihood-ratio test (Section~5.4) in the initial 
submission. The main point we were trying to make in Section~5 was that the understanding of
transformation forests as parametric models allows such procedures to be
implemented. The model-based bootstrap for the approximation of the null
distribution of likelihood-ratio statistics was introduced already in the
late 1980ies (two references were added). Its size of course depends on how
well the null model describes the data and how much the transformation
forest overfits the data. We investigate the latter issue in the new artificial
simulation experiments in Section~7, also in the presence of non-informative
predictor variables. We now explicitly mention (in the discussion) that an
empirical evaluation of likelihood-based variable importances for variable
selection, the model-based bootstrap for variability assessment and of the
likelihood-ratio test are ongoing research projects.

\end{enumerate}

\subsubsection*{Reviewer 1}

\begin{enumerate}

\item \textit{Your general framework is very clear, it is a good idea, and it fits very
              well into results which already exist about random forest.}

Thank you!

\item \label{rev1p2} \textit{You are splitting the space in such a way that the second
              expression on page 8 is maximized.}

We regret the confusion caused by the second formula on page 8, which
suggested that an exhaustive search is performed. Unbiased recursive
partitioning avoids exhaustive evaluations of all potential splits by a
separation of variable and split selection. The formula was understood
as an illustration of the concept, not the actual implementation. We
moved this formula to Section~8 and added 
a much more detailed description of variable and
split selection in transformation trees. 

\textit{If I understand that
              correctly, then you approximate the true likelihood by the
              ``exact continuous'' approximation instead of the correct
              likelihood function as on page 6.  I can see that this can
              make sense when the partitions are very small.  But isn't that
              a problem in the beginning when the partition is very big or
              in high dimensions?  Is there a computational reason for doing
              this?  Have you done simulations, how that choice affects the
              performance of your estimator?}

We are sorry for the insufficient description of the different
likelihood contributions. The ``approximation'' of the ``exact'' 
likelihood by the density is a rather uncommon (at least outside survival
analysis) concept. However, the theory of transformation models is closely
tied to the understanding and evaluation of the ``exact'' likelihood for
interval-censored observations. Because we can never observe real numbers
(only intervals), we felt it would help to think of the ``exact'' 
likelihood as a probability (the definition by Fisher dating back to 1922 in
fact advocates this point of view). We rephrased this paragraph and 
separately introduce the likelihood for real numbers and intervals now.
The new Section~8 also introduces a potential application of the
interval-censored likelihood as a means for reducing runtimes of the
algorithm.

\item \textit{In general, I wonder to what extend the argmax on page 8 is
realizable?  That seems quite costly to me.  The best split in random forest
can be found in O(p * n * log(n)).  How does this expression compare, if you
only consider axis-aligned slits of p features with n possible splitting
points?  It would be nice to see an expression of the computational
complexity and a runtime comparison in terms of n and p.}

You are right, of course. We discuss theoretical and empirical runtimes in
more detail in our new Section~8.

\item \textit{Furthermore, the score function is used as a pruning criteria
which totally makes sense for single regression trees, but in the standard
Random Forest, people usually do not do such pruning.  Have you considered
not to implement this pruning criterion?  One could perform splits until at
least k observations are left in each leaf.  Wouldn't that speed up the
simulation?  Your random forest version could perform better.}

We apologise for not being more precise in the description of our implementation
of transformation forests. Of course, forests aggregate over trees built without
internal stopping. For transformation trees it makes sense to restrict
terminal nodes to a certain number of observations because it won't be
possible to estimate the parameters of the transformation model (and thus to
compute the score matrix) with $N$ being too small. Because the size of a
terminal node cannot be controlled directly in the \textbf{randomForest}
package (the parameter \texttt{nodesize} corresponds to \texttt{minsplit}
but not \texttt{minbucket} in \textbf{rpart} or \textbf{partykit}
terminology), we required at least $25$ observations in order to implement a
split for all forest variants under test.

\item \textit{For the parametric bootstrap, why would you throw away those
``extreme'' samples?  That can make sense, but throwing them away would not be
the parametric bootstrap, and there is some intuition which I might be
missing.}

You are right, the distribution of the likelihood ratio statistic is not the
only bootstrap distribution one could be interested in. We replaced this
paragraph explaining that one can look at the bootstrap distribution of the
parameters $\parm(\rx)$ or any functional thereof (including the LR statistic).

\item \textit{I wonder whether this is the right Likelihood-Ratio Test. 
What if the true distribution is independent of $\rX$, but it is very
complicated and cannot be approximated by your unconditional maximum
likelihood model.  Then your original statistics has a completely different
distribution than your simulated H0 distribution, and you might falsely
reject $H_0$, right? }

Of course. We clarified (in Section~2, see also point ``p.6, l.-4'' by
referee 2) that we assume that the true transformation function can be
written as $\h = \basisy^\top \parm$ throughout the
manuscript. Under this condition, the procedure is just the simple
parametric bootstrap. See also point \ref{AE:boot} by the editor.

\item \textit{Furthermore, it would also be nice to have an intuitive
statement of how the choice of the set in which $\h$ lies affects the
bias-variance trade-off.}

We explicitly cover this point in the empirical evaluations in Section~7
now. The empirical effects of using the `correct' transformation function
(for example, a linear function for conditionally normal targets) compared
to an overparameterised (ie, non-linear) version are investigated
in a simulation model with normal targets. For log-normal targets, the loss
of using an underparameterised (linear) transformation function is also
presented.

\item \label{R1sim} \textit{The way I understand your simulation setup is the following:
  You take real data sets, you fit each method, and then you use the
  parametric bootstrap to create from each fitted estimator new data sets. 
  Those new data sets have the same feature points but $\rY$ values created from
  the fitted models.  To compare with fully non-parametric estimators (such as
  quantile regression forest) wouldn't it make more sense to use a bigger data
  set, use on one part of it as a training data set and use the rest as an
  evaluation data set?  At least for confidence intervals comparisons that
  would be very appealing to me and it would be fair since all of your models
  are essentially based on transformation models, right? If you need access
  to the true underlying data generating process, it might be worthwhile also
  comparing your methods to made up data generating processes.  Otherwise, you
  are comparing four estimators with QRF by creating data directly coming from
  models which are generated by those estimators.}

Our intention was to evaluate the methods based on ``realistic'' simulation
models related to a specific algorithm.  We think that our comparison with
quantile regression forests were correct, because we also directly sampled
(nonparametrically) from this model.  The main problem with a
bootstrap-based simulation is that the ``simulation model'' is a tree or a
forest and only the former can be visualised and directly understood.  We
also understand the points you and the other reviewer raised and therefore
implemented a more traditional simulation study based on artificial data
generating processes with separate learning and validation sets in
Section~7. 

\item \textit{There is also an
estimator called Density Forest and Manifold Forest, which were for example
mentioned in Criminisi et al.~2011.  It would be nice to have a comparison
to those methods as well.}

Thank you very much for pointing us to this interesting publication. Our
understanding of \url{http://dx.doi.org/10.1561/0600000035} (page 134) is that the
model is conditional normal with predictor-dependent mean and variance.
However, it is unclear (in this and other publications from the group) how
the underlying trees handle variance heterogeneity that can be explained by
the predictor variables (no details regarding the ``weak learner'' are
presented on page 137). The more flexible (as it seems) 
density and manifold forests are for the unsupervised case only.

The authors provide software for reproducing results presented in a
follow-up book (``Sherwood'' from \url{https://www.microsoft.com/en-us/download/details.aspx?id=52340&751be11f-ede8-5a0c-058c-2ee190a24fa6=True}). 

\begin{figure}
\begin{Schunk}
\begin{Sinput}
> set.seed(290875)
> N <- 1000
> x <- runif(N, min = -1, max = 1)
> y <- rnorm(N, mean = sin(2 * x * pi), sd = exp(x))
> plot(y ~ x)
> write.table(data.frame(x = x, y = y), file = "th.txt", 
+             col.names = FALSE, row.names = FALSE, sep = "\t")
\end{Sinput}
\end{Schunk}
\includegraphics{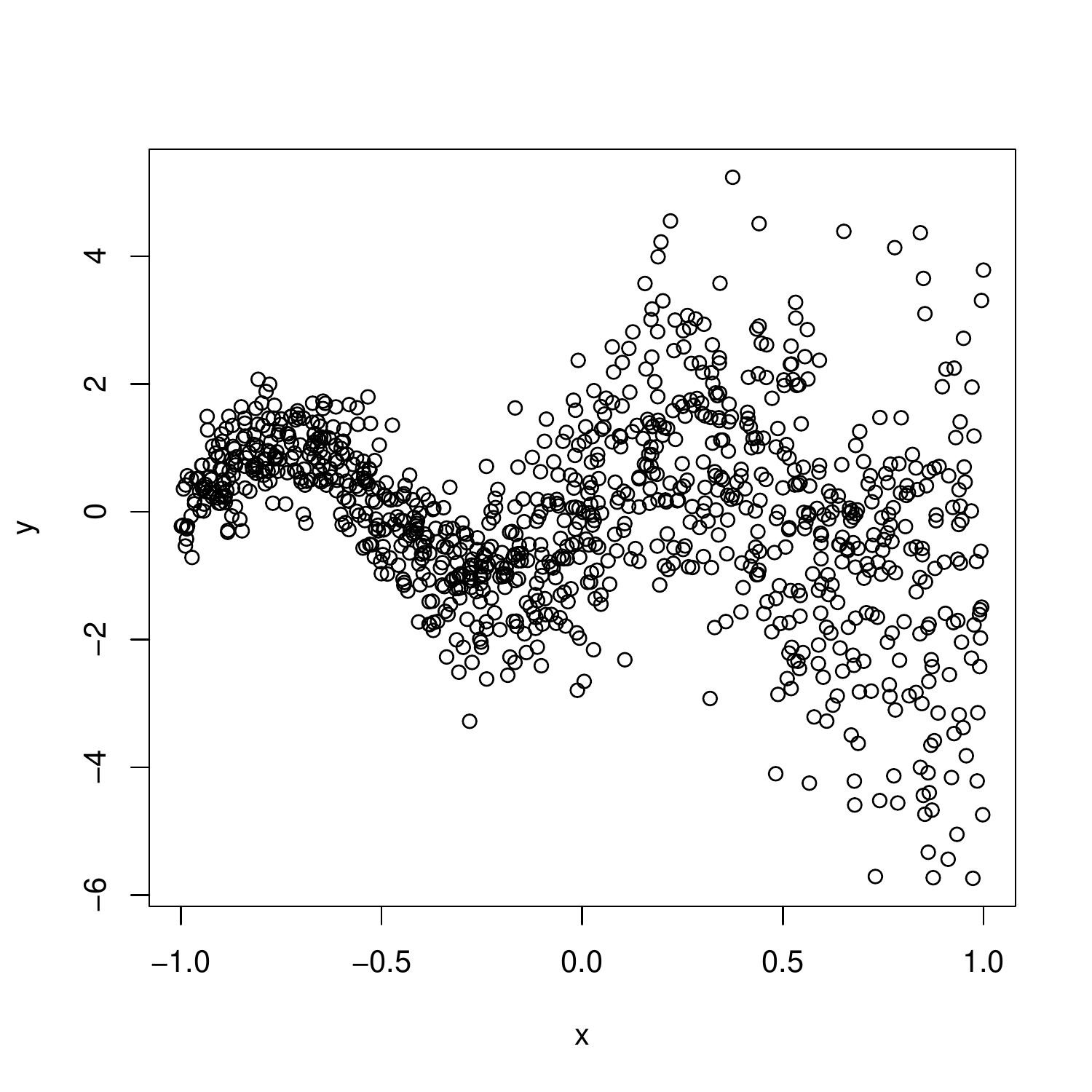}
\caption{Example to be used for ``Sherwood'' example code. \label{ex}}
\end{figure}
We generated an example data set (Figure~\ref{ex}) and tried to run the software on this example using
\begin{verbatim}
./sw regression /t 100 /d 10 th.txt
\end{verbatim}
and obtained the following output
\begin{verbatim}
Training the forest...
Trained 100 trees.         

Applying the forest to test data...
Applied 100 trees.        
Segmentation fault (core dumped)
\end{verbatim}
It also seems that the only output the software can produce is a bitmap file
with the original data overlayed with the model. We did not find any 
possibility to specify test data, let alone a means to compute out-of-sample
prediction errors etc. For these reasons it was not possible for us to
include this implementation as an additional competitor in the empirical
evaluations in Section~7. However, the method is interesting because
mean aggregation of conditional densities is performed, and this is similar
to the mean aggregation of cumulative hazard functions 
in random survival forests. The introduction now highlights this similarity.

\item \textit{For you confidence interval simulations, there are two things
to consider.  They should have the correct coverage, and they should be
small.  I would also be curious how your confidence intervals differ in
size.}

It is in general hard to compare prediction intervals, because they
condition on a specific set of predictor variables. Thus, coverage and
lengths can differ substantially over the sample space. 
We elaborate on this issue in Section~7 and also point to the
proper scoring rules literature. However, the check risk for the $10\%$ and
$90\%$ quantiles is appropriate to compare the ability of methods to
estimate these conditional quantiles and we, in addition to the negative
log-likelihood and the absolute error, report these two performance
measures. 

\end{enumerate}

\textit{Little typos: (omitted)}. Thank you, we silently corrected the
typos.

\subsubsection*{Reviewer 2}

\paragraph{Overview}

\noindent
\textit{
This is an interesting paper. Hothorn and Zeileis describe using weights, say
$w_i$ derived
from a random forest algorithm to fit the model:
\begin{eqnarray} 
\hat{\parm}^N(\rx) := \argmax_{\parm \in \Theta} \sum_{i = 1}^N w_i^N(\rx) \ell_i(\parm).
\end{eqnarray}
The likelihoods are chosen from a transformation family: $\pY(\ry) = \pZ(\h(\ry))$
Here $\pZ$ is known a priori (for example it might be the standard Gaussian distribution
function) and $\h$ is the quantity maximized. Some structure is imposed on
$\h$ to make the maximization easier. In the examples considered $\h$ was modelled
by Bernstein polynomials.}

\paragraph{Major comments}

\textit{We suggest that implementing one and only one of the major comments addressed below
would make this a very strong paper.}

Thank you for the suggestion, we chose the second option.

\begin{description} 
\item[2.1.  Developing directly relevant theoretical results.] 
  \textit{The authors cite the prior work of Hothorn, Kneib and
  B\"uhlmann showing consistency results for conditional transformation
  forests and it seems straightforward to extend these results to the case
  where the tree partition is not estimated from the data.  They go on to cite
  recent theoretical work including Scornet at al (2015) that establishes
  consistency results for an algorithm that is almost the same as Breiman's
  original random forests. They suggest that these results can be adapted to
  show consistent estimation of the likelihood function they seek to
  maximize.  However the development seems slightly heuristic and it would
  seem that some conditions on the complexity of the likelihood family is
  required but no conditions are given.  Moreover, this is not quite the
  flavour of theoretical results that are directly relevant.  We want a result
  that says that transformation forests consistently estimate the distribution
  of $\rY \mid \rX$, where convergence is measured in, for example, KS
  distance.  This is a stronger result than they develop because if, for
  example, the family of functions chosen to model $\h$ does not include the
  true $\h$ then consistent estimation of the likelihood need not imply
  consistent estimation of $\rY \mid \rX$.}

We agree that such results would be extremely important and valuable. Prior
to submitting the initial version of this manuscript, we discussed
the possibility to obtain such results with Nicolai Meins\-hausen. The outcome
of the discussion was that this task is a major research project in its own
right and we therefore refrained from digging deeper here. In addition, our
understanding of the theoretical literature on random forests is that
practically relevant results (ie, the analysis of non-idealised versions of
random forests) are very rare and technically challenging. The main
difference between Breiman and Cutler's randomForests
and the transformation forests proposed in this manuscript is with respect
to the variable and split selection. We are not aware of a paper analysing the
impact of different forms of variable and split selection on the theoretical
performance of random forests.

\item[2.2.  Providing richer simulations covering high dimensional settings.] 
  \textit{Alternatively if theoretical results are unduly difficult
  it would be worthwhile to provider a richer set of simulations.  All
  datasets from which the simulations are derived have $N \le 4177$ and $p \le
  18$; the maximal aspect ratio of the data is 16.9.  These simulation
  settings appear to be derived from Meinshausen's quantile forests paper. None
  of these settings are especially relevant for high dimensional problems,
  where Breiman's random forest has been especially successful.  While it is
  certainly true that distribution estimation requires much more data than the
  estimation of the conditional mean it seems worthwhile to at least
  investigate the performance of transformation forests in high dimensional
  settings.}

We agree with your criticism and the criticism raised by reviewer 1 (see
\ref{R1sim} by reviewer 1). In the novel simulations based on artificial
data generating processes we study the performance of all methods in the
low-dimensional situation and the situation with $50$ noise variables added.

\end{description}

\paragraph{Minor comments}

\begin{description}
\item[p.2, l.15] 
  \textit{The use of random survival forests is preceded by the paper ``Tree
  structured survival analysis'' (1985) by Gordon and Olshen and arguably
  should be cited, although, because it predates the random forest algorithm,
  they adapt trees rather than forests.}

You are right, of course. In this paragraph, we focus on existing random
forest methods for the estimation of conditional distribution (or survivor)
functions. We incorrectly cited random survival forests here, because
aggregation takes place by averaging cumulative hazard functions (their
formula 3.2), so this reference was removed. 

\item[p.2, l.16] \textit{Lin and Jeon (2006) certainly develop ``adaptive
  nearest neighbours'' but it is not obvious to me that they propose they be
  used for conditional distribution estimation as stated.}

Yes, they exclusively focus on conditional means. We highlight this fact now
prominently.

\item[p.6, l.12] \textit{The fact that $\pZ (\infty) = 1$ and $\pZ(-\infty) = 0$ 
  is implied by the fact that $\pZ$ is a distribution function and should
  not be stated.}

Sorry, a copy-and-paste error from the (more general) description in
\url{http://dx.doi.org/10.1111/sjos.12291}. Fixed.

\item[p.6, l.16] \textit{Often the minimum extreme value distribution is
  parameterized by the mean and variance to which the minimum value
  corresponds; consider calling it the standard minimum extreme value
  distribution.}

Thank you, fixed.

\item[p.6, l.-8] \textit{``relatively precise measurements''.  You seem
  just to need that the density be relatively constant within an interval; the
  interval need not be too short.}

See \ref{rev1p2} by Referee 1.

\item[p.6, l.-4] \textit{You parameterize $\h$ to ``simplify estimation''. 
  I think this is required for more reasons than simplicity.  If $\h$ were
  allowed to be any function then there would be no hope of estimating it.}

You are right. Throughout the manuscript we assume that the true unknown
transformation function can be written in terms of such a basis function. We
clarified this fact.

\item[p.8, l.10] \textit{Unbiasedness seems to be used in the sense that
  variables with more classes are not preferred for splits in the null model
  of no effect.  This is not quite obvious from the context so a sentence
  clarifying the use of the term would be helpful.}

We added a more elaborate description of the variable and split
selection procedure, including a discussion of ``unbiasedness'', to our new
Section~8.

\item[p.9, s5.2] \textit{Permutation variable importance is used for random
  forests despite its many drawbacks because better alternatives are harder to
  come by.  (For example if two features are equal but highly correlated with
  $\rY$ they will seem to have zero importance by this measure).  Given the
  richer parametric structure your estimating it would be nice if a more
  satisfying measure of variable importance were available.}

We agree that permutation variable importance is not an ideal tool for
variable selection. We elaborate on the possibility of a 
conditional variable importance in the revision and state that more
empirical insight into the variable selection properties are necessary. We
are not convinced that the parametric structure gives us more leverage,
because ``only'' the conditional parameter functions $\parm(\rx)$, \ie
random forest-type black-box functions, determine variable importance in
transformation forests.

\item[p.9, s5.3] \textit{Can you provide any theoretical guarantees for the
  model based bootstrap?  Typically in low dimensions guarantees are not too
  difficult to obtain.  It would also be deduce whether it fails in high
  dimensions but this may be quite arduous.}

We added (in Section~7) some empirical evidence on the overfitting behaviour
of transformation forests. It seems that the amount of overfitting,
especially under the null of no association between predictors and target,
seems tolerable, indicating that the likelihood-ratio test and also the
model-based bootstrap might perform appropriately. However, we did not yet
perform tailored experiments on these questions (and explicitly point the
reader to ongoing research on these topics in the discussion).

\item[p.11,l13] \textit{Consider also citing Breiman (2004) ``Consistency for
  a simple model of random forests''.  This is the earliest argument of which I
  know for the consistency of random forests.}

Thank you!

\item[general] \textit{Consider citing ``Piecewise-polynomial regression trees''
  (1994).}

SUPPORT is one of the earlier references for fitting model-based trees. We
reviewed these predecessors of model-based recursive partitioning in
Zeileis, Hothorn and Hornik (2008) and refer to this review instead. 

\item[general] \textit{Consider citing ``Supervised neighborhood for
  distributed nonparametric regression'' by Adam Bloniarz, Christopher Wu, Bin
  Yu and Ameet Talwalker (2016).  They exploit the neighborhood of Lin and
  Jeon (2006) in a similar fashion but for a different purpose.}

Thank you for pointing us to this relevant paper on local linear model
estimation. The paper is now cited in the Introduction and Section~2. 

\end{description}

\end{appendix}

\end{document}